\documentclass[aps,prb,twocolumn,superscriptaddress,nopacs,nofootinbib]{revtex4}

\usepackage{amsmath,amssymb,latexsym}
\usepackage{graphicx}
\graphicspath{{../fig/}}
\usepackage{txfonts}
\usepackage{mathrsfs}
\usepackage{color}
\DeclareSymbolFont{symbols}{OMS}{cmsy}{m}{n}
\DeclareSymbolFont{largesymbols}{OMX}{cmex}{m}{n}
\newcommand{\bm}[1]{\boldsymbol #1}

\begin{document}

\title{
Nonlinear light-Higgs coupling in superconductors beyond BCS: \\
Effects of the retarded phonon-mediated interaction
%driven by ac electric fields
}

\author{Naoto Tsuji}
\affiliation{RIKEN Center for Emergent Matter Science (CEMS), Wako 351-0198, Japan}
\author{Yuta Murakami}
\affiliation{Department of Physics, University of Tokyo, Hongo, Tokyo 113-0033 , Japan}
\affiliation{Department of Physics, University of Fribourg, 1700 Fribourg, Switzerland}
\author{Hideo Aoki}
\affiliation{Department of Physics, University of Tokyo, Hongo, Tokyo 113-0033 , Japan}
\affiliation{Electronics and Photonics Research Institute, Advanced Industrial Science and Technology (AIST), Tsukuba, Ibaraki 305-8568, Japan}
%\email[]{}
%\homepage[]{}
%\thanks{}
%\altaffiliation{}

\begin{abstract}
We study the contribution of the Higgs amplitude mode on the nonlinear optical response of superconductors 
beyond the BCS approximation by taking into account the retardation effect 
in the phonon-mediated attractive interaction. To evaluate the vertex correction in nonlinear optical susceptibilities
that contains the effect of collective modes, we propose an efficient scheme which we call the ``dotted DMFT''
based on the nonequilibrium dynamical mean-field theory (nonequilibrium DMFT) to go around the difficulty of
solving the Bethe-Salpeter equation and analytical continuation.  
The vertex correction is represented by the derivative of the self-energy with respect to the external driving field, 
which is self-consistently determined by the differentiated (``dotted'') DMFT equations.
We apply the method to the Holstein model, a prototypical electron-phonon-coupled system,
to calculate the susceptibility for the third-harmonic generation including the vertex correction.
The results show that, in sharp contrast to the BCS theory, 
the Higgs mode can contribute to the third-harmonic generation 
for general polarization of the laser field
with an order of magnitude comparable to the contribution from the pair breaking or charge density fluctuations.
The physical origin is traced back to the nonlinear resonant
light-Higgs coupling, which has been absent in the BCS approximation.
\end{abstract}

%\collaboration{}
%\noaffiliation

\date{\today}

\pacs{
%74.25.N-, 74.40.Gh, 71.10.Fd
%74: Superconductivity
%74.25.N-: Response to electromagnetic fields
%74.40.Gh: Nonequilibrium superconductivity
%71.10.Fd: Lattice fermion models (Hubbard model, etc.)
}

\maketitle

\section{Introduction}
\label{introduction}

Nonequilibrium dynamics of superconductors induced by intense laser excitations
%with relatively low frequency such as in the terahertz (THz) and mid-infrared regime
opens various possibilities of controlling emergent states of matter without destroying quantum coherence.
\cite{Fausti2011,Cortes2011,Smallwood2012,DalConte2012,Kaiser2012,Matsunaga2012,Hinton2013,Mansart2013,Matsunaga2013,
Hu2014,Mankowsky2014,Matsunaga2014,Mitrano2016,Giannetti2016}
Specifically, for relatively low frequencies of the laser such as terahertz (THz) and mid-infrared,
we can expect to suppress the generation of quasiparticles having high energies
that might be quickly transformed into heat through inelastic collisions causing a destruction of quantum coherence.
Recent experiments indeed report that a superconducting-like state can be generated from the normal state by 
such low-energy excitations.\cite{Fausti2011,Kaiser2012,Hu2014,Mankowsky2014,Mitrano2016}

In superconductors, there exists a collective mode called the Higgs amplitude mode,
which plays an important role in low-energy dynamics. The mode corresponds to the coherent amplitude oscillation
of the superfluid density, which has a long history of theoretical studies.
\cite{Anderson1958,NambuJonaLasinio1961,Anderson1963,EnglertBrout1964,Higgs1964,Guralnik1964,VolkovKogan1973,
LittlewoodVarma1981a,Kulik1981,Varma2002,BarankovLevitovSpivak2004,YuzbashyanTsyplyatyevAltshuler2006,
BarankovLevitov2006,YuzbashyanDzero2006,
Papenkort2007,Papenkort2008,Gurarie2009,Schnyder2011,Podolsky2011,
VolovikZubkov,TsujiEcksteinWerner2013,BarlasVarma2013,Tsuchiya2013,Gazit2013,Krull2014,Cea2014,PekkerVarma2015,Kemper2015,Cea2015,Tsuji2015,Peronaci2015,Jujo2015,Krull2015,
Murakami2016a,Cea2016,Murotani2016,Murakami2016b}
Experimental observation of the Higgs mode in superconductors has been reported by Raman scattering,
\cite{SooryakumarKlein1980,Measson2014}
and THz pump-probe experiments.\cite{Matsunaga2013,Matsunaga2014}
It has also been reported in a THz pump experiment\cite{Matsunaga2014} that 
there emerges a third-harmonic generation (THG)
in the nonlinear optical response that is resonantly enhanced when the doubled frequency ($2\Omega$)
of the incident light
equals the superconducting gap ($2\Delta$), which coincides with the energy of the collective Higgs mode
at long wavelength. On the other hand, there also exist individual excitations
(Cooper pair breaking or charge density fluctuations), whose lower bound in the energy spectrum
resides at the same energy of $2\Delta$
with a diverging density of states. The question then is to what extent these two contribute to
the nonlinear optical response in superconductors and how strongly the light is nonlinearly coupled
to the Higgs mode.\cite{Tsuji2015,Cea2016}

In the BCS mean-field theory 
(with the random phase approximation),
the contribution of pair breaking or charge density fluctuation
to the THG susceptibility is expressed in a gauge invariant form (including the screening effect) as\cite{Cea2016}
\begin{align}
\chi_0^{\rm BCS}(\Omega)
&=
\sum_{\bm k} (\ddot\epsilon_{\bm k})^2 \chi_{33}(\bm k,\Omega)
-\frac{\left[\sum_{\bm k} \ddot\epsilon_{\bm k}\chi_{\rm 33}(\bm k,\Omega)\right]^2}{\sum_{\bm k}\chi_{\rm 33}(\bm k,\Omega)}.
\label{chi BCS}
\end{align}
Here $\epsilon_{\bm k}$ is the band dispersion, $\ddot\epsilon_{\bm k}=\sum_{ij} (\partial^2\epsilon_{\bm k}/\partial k_i \partial k_j)
e_i e_j$, $\bm e$ is the polarization vector of light, and
\begin{align}
\chi_{33}(\bm k,\Omega)
&=
-\frac{i}{2}\int \frac{d\omega}{2\pi} {\rm Tr}[\tau_3 \hat G_{\bm k}(\omega+2\Omega)\tau_3 \hat G_{\bm k}(\omega)]^<,
\end{align}
where $\tau_3$ is the third component of the Pauli matrix,
$\hat G_{\bm k}(\omega)$ is the Nambu-Gor'kov Green's function, and $<$ denotes the lesser component
based on the Langreth rule\cite{Langreth1976}
(with the notation defined in Appendix \ref{notation}).
For $s$-wave superconductors, $\hat G_{\bm k}(\omega)$ and $\chi_{33}(\bm k,\omega)$
depend respectively on the momentum through $\epsilon_{\bm k}$, which allows one to change the momentum sum
into an energy integral by inserting $1=\int d\epsilon \delta(\epsilon-\epsilon_{\bm k})$. 
Then we can define expansions around the Fermi energy,\cite{Tsuji2015,Murotani2016}
\begin{align}
\sum_{\bm k} \delta(\epsilon-\epsilon_{\bm k}) \ddot\epsilon_{\bm k}
&=
D(\epsilon_F)(c_0+c_1 \epsilon+c_2 \epsilon^2+\cdots),
\label{expansion1}
\\
\sum_{\bm k} \delta(\epsilon-\epsilon_{\bm k}) (\ddot\epsilon_{\bm k})^2
&=
D(\epsilon_F)
(\tilde c_0+\tilde c_1 \epsilon+\tilde c_2 \epsilon^2+\cdots),
\label{expansion2}
\end{align}
where $D(\epsilon_F)$ is the density of states at the Fermi energy.
%In the low-energy limit, $(\ddot\epsilon_{\bm k})^2$ and $\ddot\epsilon_{\bm k}$ in Eq.~(\ref{chi BCS})
%can be replaced with $D(\epsilon_F)\tilde c_0$ and $D(\epsilon_F)c_0$, respectively.
In Ref.~\onlinecite{Tsuji2015}, it is assumed that the constant terms in the expansions (\ref{expansion1}) and (\ref{expansion2})
can be removed by gauge transformations,
so that the pair breaking effect in THG is less dominant than the Higgs mode.
As pointed out in Ref.~\onlinecite{Cea2016}, however, this holds
in rather restricted situations, such as one dimensional (1D) lattices, 2D square and 3D simple cubic lattices with polarization $\bm e$
respectively parallel to $(1,1)$ and $(1,1,1)$ directions, 3D body-centered-cubic lattice with $\bm e$ parallel to $(1,0,0)$, and so on.
For a general lattice with a general polarization, the constant terms may survive, and the Higgs-mode contribution
may be left subleading.

\begin{figure}[t]
\begin{center}
\includegraphics[width=6cm]{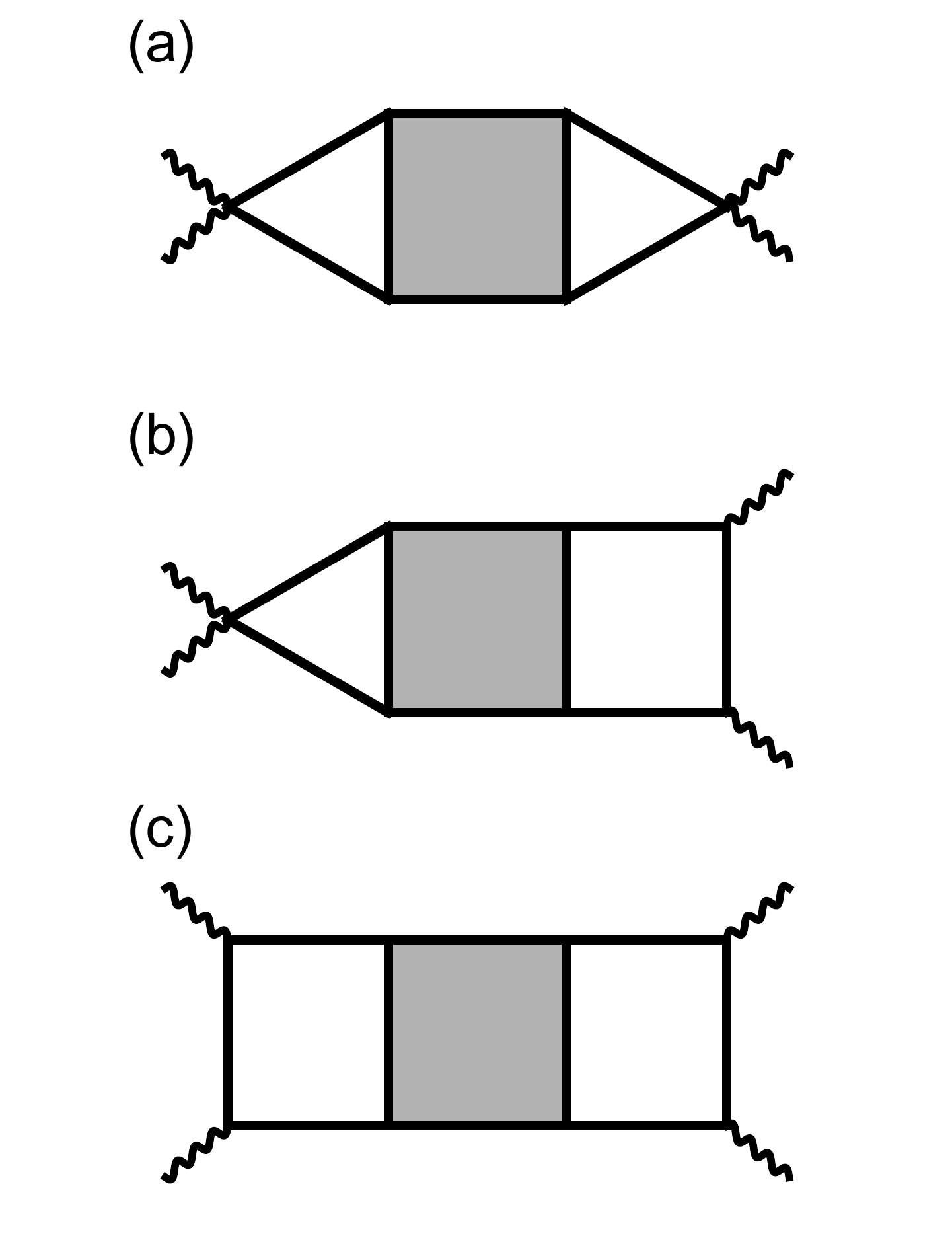}
\caption{
Feyman diagrams for the non-resonant (a), mixed (b), and resonant (c) 
contributions to the THG susceptibility containing the effect of collective modes
as vertex corrections. The solid (wavy) lines represent the electron (photon) propagators,
while the shaded boxes represent the reducible four-point vertex function. Among the four photon lines,
one is outgoing with an energy $3\Omega$, and the other three are incoming with an energy $\Omega$.}
\label{raman diagram}
\end{center}
\end{figure}

Then the next question is: what will happen if one goes beyond the BCS approximation. 
In fact, the superconductor NbN used in the experiments\cite{Matsunaga2013,Matsunaga2014}
is known to have a strong electron-phonon coupling ($\lambda\sim 1$),
\cite{Kihlstrom1985,Brorson1990,Chockalingam2008}
where it is important to capture corrections from the BCS analysis.
%which causes the retarded interaction.
Indeed the argument in the previous paragraph heavily relies on
the speciality of BCS: the (nonlinear) coupling to the light occurs only in a non-resonant form
$\ddot\epsilon_{\bm k}A(t)^2$ rather than in a resonant form $\dot\epsilon_{\bm k}A(t)\dot\epsilon_{\bm k}A(t')$,
where $\bm A(t)=A(t)\bm e$ is the vector potential, and $\dot\epsilon_{\bm k}=\sum_i (\partial\epsilon_{\bm k}/\partial k_i) e_i$. 
The terminology (``resonant'' and ``non-resonant'') is here borrowed 
from literatures on Raman scattering.\cite{Devereaux2007}
These forms can be expressed as diagrams for the THG susceptibility
\cite{ButcherCotter} in Fig.~\ref{raman diagram}
(which in fact very much resemble Raman-scattering diagrams\cite{Devereaux2007}),
where the effect of collective modes is incorporated in the vertex correction, with the Higgs mode represented by
an infinite series of ring diagrams in the $\tau_1$ channel.\cite{LittlewoodVarma1981a,Varma2002,PekkerVarma2015,Tsuji2015}
Two photon lines attached together to electron lines represent the non-resonant coupling,
while two single-photon lines attached separately represent the resonant coupling.

Within the BCS theory, 
there is only the non-resonant coupling\footnote{This can be understood in Anderson's pseudospin picture.\cite{Anderson1958,Matsunaga2014,Tsuji2015}
The time-dependent BCS theory is equivalent to a pseudospin dynamics described by $\partial\bm\sigma_{\bm k}/\partial t
=2\bm b_{\bm k}\times\bm \sigma_{\bm k}$, where $\bm\sigma_{\bm k}$ is the pseudospin, and
$\bm b_{\bm k}=(-{\rm Re}\Delta,-{\rm Im}\Delta,(\epsilon_{\bm k+\bm A(t)}+\epsilon_{\bm k-\bm A(t)})/2)$
is the pseudomagnetic field. The coupling to the light is provided by the $z$ component of the pseudomagnetic field,
$\epsilon_{\bm k}+\ddot\epsilon_{\bm k}A(t)^2/2+O(A^4)$, which is in a form of the non-resonant coupling.},
and the mixed [Fig.~\ref{raman diagram}(b)] and resonant (c) contributions to THG exactly vanish. 
This is confirmed by explicitly calculating 
the convolution of relevant three electron propagators,
\begin{align}
\int \frac{d\omega}{2\pi} {\rm Tr}
[\tau_1 \hat G_{\bm k}(\omega+2\Omega) \hat G_{\bm k}(\omega+\Omega) \hat G_{\bm k}(\omega)]^<
&=0
\quad (\mbox{BCS}).
\label{GGG BCS}
\end{align}
However, this does not guarantee that these contributions would remain small
if one goes beyond the BCS approximation. For example, the real part of the optical conductivity $\sigma(\Omega)$
vanishes for $\Omega\neq 0$ within the BCS theory, since
\begin{align}
\int \frac{d\omega}{2\pi} {\rm Tr}
[\hat G_{\bm k}(\omega+\Omega)\hat G_{\bm k}(\omega)]^<
&=0
\quad (\mbox{BCS}),
\end{align}
in much the same way as in Eq.~(\ref{GGG BCS}). In reality, however,
the real part of the optical conductivity is nonzero and not even small.\cite{Matsunaga2013,Zimmermann1991}
They become nonzero when one takes account of dynamical correlations such as
the electron-phonon coupling (producing retarded interactions) or impurity scattering.
In those situations, we can expect that the resonant and mixed contributions to the THG response may also be nonzero.
Indeed, it has been shown in the study of Raman scattering for correlated electron systems that the resonant contribution can
significantly enhance the non-resonant Raman response.\cite{Shvaika2004,Shvaika2005}

This has motivated us to study here the nonlinear optical response of superconductors
for electron-phonon coupled systems beyond the BCS approximation. Theoretically,
it is quite challenging to evaluate all of the non-resonant, mixed and resonant diagrams involving
the four-point vertex on an equal footing, since the vertex carries three independent momenta and frequencies.
Therefore, we employ the dynamical mean-field theory (DMFT),\cite{GeorgesKotliarKrauthRozenberg1996}
which assumes the momentum-independent self-energy and vertex function. Still, the calculation
is quite demanding 
if one tries to evaluate the nonlinear response function
by solving the Bethe-Salpeter equation and performing multiple analytical continuations.
In higher dimensions in the thermodynamic limit,
%(which can accommodate the superconducting phase),
an analysis including the vertex correction has so far been performed
only in exceptional cases, such as the Raman response of the Falicov-Kimball model.
\cite{Shvaika2004,Shvaika2005,FreericksZlatic2003,Matveev2010}
For the Hubbard model, the nonlinear optical response has been analyzed 
by Hartree-Fock approximation,\cite{Jujo2006}
by DMFT without considering vertex corrections,\cite{Jujo2008,Jujo2009}
and by exact diagonalization for small finite-size systems.\cite{Mizuno2000,Takahashi2002}
For the Holstein model, higher-harmonic generation has been studied 
by Migdal approximation without considering vertex corrections.\cite{Kemper2013}
For 1D Hubbard-Holstein model, THG response has been studied by 
the density-matrix renormalization group.\cite{Sota2015}

In this paper, we propose an efficient way to calculate the vertex correction
for nonlinear optical susceptibilities, which we call the ``dotted DMFT''\cite{TsujiPhD}, without
directly solving the Bethe-Salpeter equation and performing analytical continuation.
The idea is to let the nonequilibrium DMFT equations\cite{noneqDMFTreview} differentiated (``dotted'') with respect to
the external field to deduce a self-consistent equation for the vertex function
represented by the dotted self-energy. We then apply the method to the Holstein model,
a prototypical model for electrons interacting with local phonons giving retarded interactions
among electrons. The results indicate
that the resonant contribution from the Higgs mode to the THG susceptibility can indeed be comparable to 
those from the pair breaking or density fluctuations.
In particular, the resonance of THG at $2\Omega=2\Delta$ can be enhanced by the nonlinear resonant coupling
between the light and Higgs mode.

The paper is organized as follows. In Sec.~\ref{model}, we describe the model set-up that we use 
throughout the paper for the analysis of the nonlinear optical response in superconductors.
In Sec.~\ref{dotted dmft}, we propose an efficient method
(dotted DMFT)
to evaluate the vertex correction for dynamical susceptibilities based on the nonequilibrium DMFT.
Sec.~\ref{results} describes the results of the THG susceptibility obtained by the dotted DMFT
for the electron-phonon-coupled system. In Sec.~\ref{summary} we summarize the paper.

\section{Model}
\label{model}

We take the Holstein model as a typical model for electrons interacting with
local phonons,
\begin{align}
H&=
\sum_{ij,\sigma}
t_{ij} (c_{i\sigma}^\dagger c_{j\sigma}+{\rm h.c.})
-\mu\sum_i n_i
\notag
\\
&\quad
+\omega_0 \sum_i b_i^\dagger b_i
+g\sum_i (b_i+b_i^\dagger)(n_i-1).
\end{align}
Here $c_{i\sigma}^\dagger$ ($c_{i\sigma}$) is the creation (annihilation) operator for an electron
at site $i$ with spin $\sigma=\uparrow,\downarrow$,
$t_{ij}$ is the hopping amplitude,
$n_i=\sum_\sigma c_{i\sigma}^\dagger c_{i\sigma}$, 
$\mu$ is the chemical potential, 
$b_i^\dagger$ ($b_i$) is the creation (annihilation) operator for phonons
having a frequency $\omega_0$, and $g$ is
the electron-phonon coupling constant. We then apply the dynamical mean-field theory (DMFT)
to solve the model. Since DMFT becomes exact for large spatial dimensions ($d\to\infty$),
\cite{MetznerVollhardt1989,GeorgesKotliarKrauthRozenberg1996} we take the 
hypercubic lattice, whose energy dispersion is
\begin{align}
\epsilon_{\bm k}&=
-2t\sum_{i=1}^d \cos k_i.
\label{ek}
\end{align}
As usually done, we scale the hopping as
$t=t^\ast/\sqrt{2d}$ with a fixed $t^\ast$ to obtain a meaningful fixed point in the large $d$ limit, 
which results in a gaussian density of states
$D(\epsilon)=e^{-\epsilon^2/2t^\ast{}^2}/\sqrt{2\pi}t^\ast$.
We use $t^\ast$ as the unit of energy (frequency) throughout the paper.
We concentrate on the half-filled electron system ($\mu=0$), in which the particle-hole symmetry is
fully respected. 
In the particle-hole symmetric case,
the Higgs amplitude mode is safely decoupled from the phase mode, and the screening effect is absent.
Away from half filling, the amplitude mode can hybridize with the phase mode in principle.
However, we expect that the damping of the amplitude mode into the phase mode is suppressed
in superconductors, since the phase mode is pushed to high energies ($\sim$ the plasma frequency)
due to the Anderson-Higgs mechanism.\cite{Anderson1963,EnglertBrout1964,Higgs1964,Guralnik1964}

If one integrates out the phonon degrees of freedom, the electrons 
acquire an effective retarded interaction,
\begin{align}
U(\omega)&=
g^2 D_0^R(\omega),
\label{U(w)}
\end{align}
where $D_0^R(\omega)$ is the noninteracting retarded phonon Green's function,
\begin{align}
D_0^R(\omega)
&=
\frac{2\omega_0}{(\omega+i\gamma)^2-\omega_0^2}.
\label{D0R}
\end{align}
We introduce a parameter $\gamma$ to regularize the phonon Green's function.
In the static limit ($\omega\to 0$), the effective interaction approaches 
$U(\omega=0)=-2g^2\omega_0/(\omega_0^2+\gamma^2)<0$, i.e., the attractive interaction.
The strength of the attractive interaction can be measured (within the unrenormalized Migdal approximation
as introduced later)
by a dimensionless parameter,
\begin{align}
\lambda&\equiv
|U(\omega=0)| D(\epsilon_F)
=
\frac{2g^2\omega_0 }{\omega_0^2+\gamma^2}D(\epsilon_F).
\label{lambda}
\end{align}
%where $D(\epsilon_F)$ is the density of states at the Fermi energy.
When the attractive interaction is large enough and the temperature is low enough, the model exhibits a phase transition
from the normal to superconducting states.

%Usually, the parameter $\gamma$ should be taken to be infinitesimal ($\gamma\to +0$).
In this paper, instead of taking the infinitesimal limit of $\gamma$ ($\to +0$),
we keep it nonzero and regard it as a phenomenological parameter
that represents the finite lifetime ($\tau\sim\gamma^{-1}$) of phonon oscillations.
%and take a nonzero $\gamma$.
This is physically natural, since the phonon oscillation should be damped to some extent in real solids 
by various possible ways of scattering and energy dissipation.
The electron-phonon coupling itself can induce the damping of phonons.\cite{Murakami2015}
A finite $\gamma$ is not necessarily phenomenological,
but can be actually modeled by phonons coupled to a heat bath comprising 
many harmonic oscillators (Caldeira-Leggett-type model\cite{Caldeira1983}).
In the application of the dotted DMFT, 
which we shall introduce in the next section, it turns out that 
it is important to take a nonzero $\gamma$ (avoiding infinitely long-lived phonons)
to stabilize the convergence of the dotted DMFT calculation.

In a similar manner, we introduce a small imaginary part $\delta$ (broadening factor) in the noninteracting retarded electron Green's function,
\begin{align}
G_{0\bm k}^R(\omega)
&=
\frac{1}{\omega+i\delta+\mu-\epsilon_{\bm k}},
\end{align}
where
$\delta$ can be considered as a decay rate of noninteracting electrons. It can be modeled by electrons
coupled to a bath composed of free fermions.\cite{noneqDMFTreview,TsujiOkaAoki2009}
Compared to $\gamma$, the stability of the dotted DMFT is less sensitive to $\delta$,
so that we can take a much smaller value for $\delta$ than for $\gamma$.

To study the third-harmonic generation, we apply an ac electric field to the Holstein model.
We use the temporal gauge to represent the electric field with a vector potential $\bm A(t)=\bm e A e^{-i\Omega t}$,
where $\bm e$ is the polarization vector ($||\bm e||=1$), $A$ and $\Omega$ is the amplitude and frequency
of the vector potential, respectively. $A$ is related to the amplitude of the electric field $E$ via $A=E/(i\Omega)$.
The ac field is minimally coupled to the electrons through the Peierls phase. The resulting form of the coupling
is $\sum_{\bm k\sigma} \epsilon_{\bm k+\bm A(t)} c_{\bm k\sigma}^\dagger c_{\bm k\sigma}$
in the kinetic term of the Hamiltonian, 
where we have put the elementary charge $e=1$.
The electron current is defined by
\begin{align}
\bm j(t)&=
-i\sum_{\bm k} \bm v_{\bm k+\bm A(t)} G_{\bm k}^<(t,t),
\label{current}
\end{align}
where $\bm v_{\bm k}=\partial\epsilon_{\bm k}/\partial {\bm k}$ is the group velocity.
%$v_{\bm k}^i=\frac{\partial\epsilon_{\bm k}}{\partial k_i}$ is the group velocity.
We measure the current along the electric field,
\begin{align}
j(t)=\bm j(t)\cdot\bm e.
\end{align}
The susceptibility for the third-harmonic generation $\chi(\Omega)$ is defined by the nonlinear current
oscillating with the frequency $3\Omega$,
\begin{align}
j^{(3)}(t)
&=
\chi(\Omega) A^3 e^{-3i\Omega t}.
\label{THG susceptibility}
\end{align}
To obtain the THG susceptibility, we take the third derivative of Eq.~(\ref{current}) with respect to $A$ and then set $A=0$.
This involves the derivatives of $G$, which are evaluated by means of the dotted DMFT, as will be explained in the next section.

Since our model is infinite dimensional, it is not obvious how to choose
the polarization. One convenient way is to take the direction parallel to $(1,1,\dots,1)$,
for which every direction is equivalent. However, as mentioned in the introduction,
this choice has a bias that the pair breaking effect is suppressed in the THG response.
Another simple choice is $(1,0,\dots,0)$. This, on the other hand, is the direction
that maximally enhances the pair breaking contribution. To let the situation as fair as possible,
we choose a general direction
\begin{align}
\bm e
\propto
(\overbrace{\underbrace{1,1,\dots,1}_m,0,\dots,0}^d),
\label{polarization}
\end{align}
where $m$ is the number of dimensions along which the polarization vector has nonzero components.
It is a kind of generalization of $(1,1,0)$ direction for the three-dimensional cubic lattice.
We fix the ratio,
\begin{align}
\alpha=\frac{m}{d}
\quad
(0\le \alpha \le 1),
\end{align}
and take the limit of $d, m\to\infty$. 
%with a fixed $\alpha\sim O(1)$.
The parameter $\alpha$ continuously interpolates the two limits of $(1,1,\dots,1)$ and $(1,0,\dots,0)$.

The advantage of this setup is that it greatly simplifies the momentum integral without putting a bias on the pair breaking effect.
We need a very fine grid for the momentum integral to eliminate the finite-size effect, which is particularly severe 
in the calculation of the THG spectrum, since one has to resolve the superconducting gap structure in the very vicinity of the Fermi energy.
One might apply DMFT to the two-dimensional square lattice as an approximation (instead of applying to the hypercubic lattice), 
but our experience indicates that the number of $k$ points that has to be taken is so huge that it is practically intractable.

\begin{figure*}[t]
\begin{center}
\includegraphics[width=15cm]{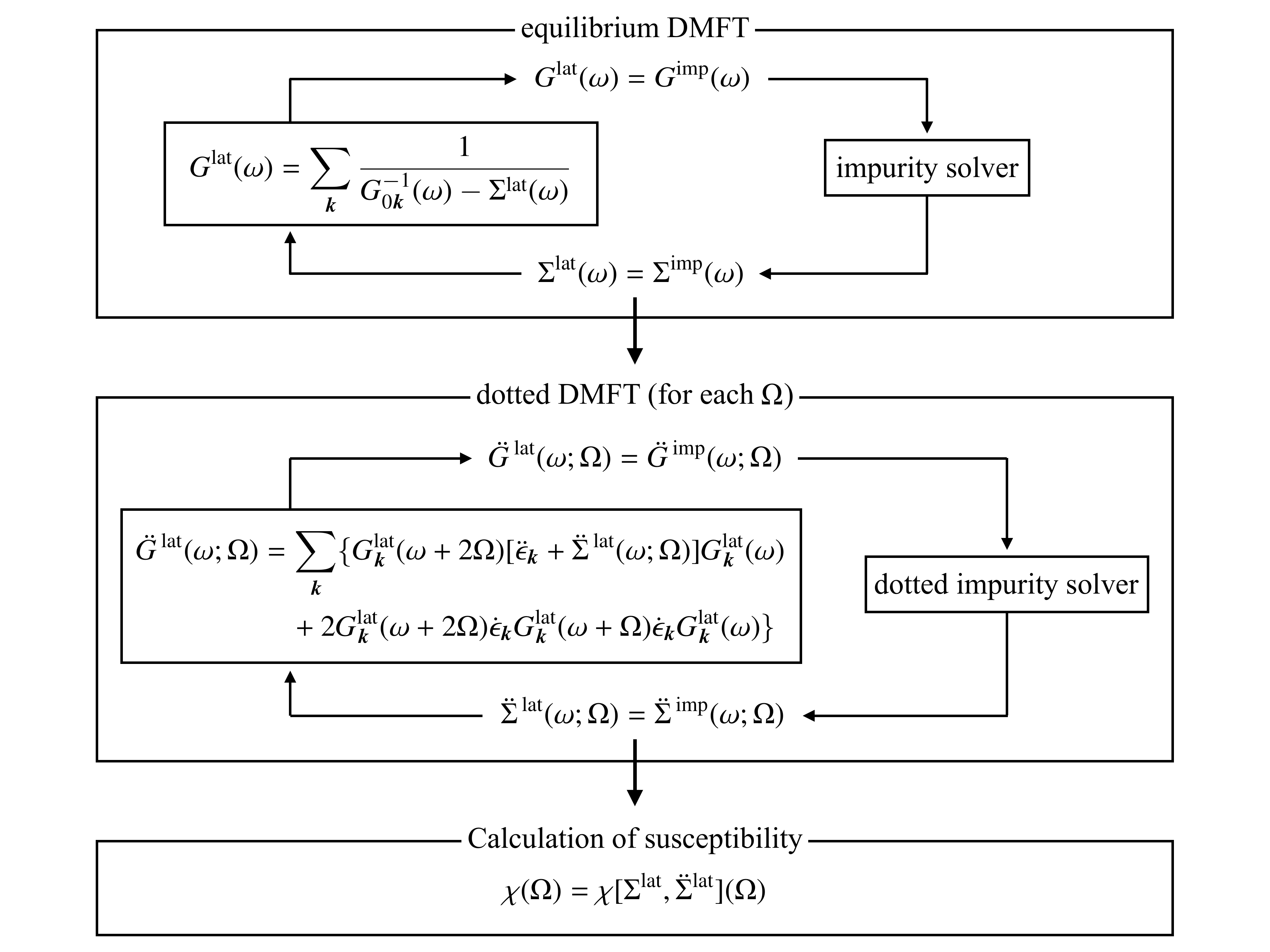}
\caption{A schematic picture for the dotted dynamical mean-field theory (dotted DMFT) formalism 
for the third-harmonic generation. Each equation holds for
the retarded, advanced, lesser, and greater Green's functions and self-energies, respectively.
%The lesser and greater components of the products of Green's functions and the self-energy should be
%understood according to the Langreth rule\cite{Langreth1976} [see Eq.~(\ref{dotted lattice Dyson2})], where the lesser and greater components of 
%$\dot\epsilon_{\bm k}$ and $\ddot\epsilon_{\bm k}$ are put to be zero. 
An analogous treatment can be generally applied to arbitrary dynamical susceptibilities.}
\label{dotted DMFT algorithm}
\end{center}
\end{figure*}

To see how the momentum integral is simplified, we expand $\epsilon_{\bm k+\bm A(t)}$ in $A$ as
\begin{align}
\epsilon_{\bm k+\bm A(t)}
&=
\epsilon_{\bm k}+\dot\epsilon_{\bm k}A e^{-i\Omega t}
+\frac{1}{2}\ddot\epsilon_{\bm k}A^2 e^{-2i\Omega t}
+\cdots,
\end{align}
where the dot denotes the derivative with respect to $A$, that is,
\begin{align}
\dot\epsilon_{\bm k}
&=
\sum_{i=1}^d \frac{\partial\epsilon_{\bm k}}{\partial k_i} e_i,
\label{dot ek}
\\
\ddot\epsilon_{\bm k}
&=
\sum_{i,j=1}^d \frac{\partial^2\epsilon_{\bm k}}{\partial k_i \partial k_j} e_i e_j,
\label{ddot ek}
\end{align}
and so on. The THG susceptibility is expressed as a momentum integral of a function of $\epsilon_{\bm k}$
multiplied by some of $\dot\epsilon_{\bm k}, \ddot\epsilon_{\bm k}, \dddot\epsilon_{\bm k}$, and $\ddddot\epsilon_{\bm k}$
[with the total number of derivatives being always four since $\bm v_{\bm k}$ in the definition of the current (\ref{current}) contains one derivative while the other three come from the external field]. 
For instance, let us consider a momentum integral of the form
\begin{align}
&\quad
\sum_{\bm k} \dot\epsilon_{\bm k} \dddot\epsilon_{\bm k} f(\epsilon_{\bm k})
\notag
\\
&=
\frac{1}{m^2}\sum_{\bm k} \sum_{i=1}^m 2t\sin k_i \sum_{j=1}^m (-2t)\sin k_j  f(\epsilon_{\bm k})
\notag
\\
&=
-\frac{4t^2}{m^2}\sum_{\bm k} \left(\sum_{i=1}^m \sin^2 k_i+\sum_{\substack{i,j=1\\ i\neq j}}^m \sin k_i \sin k_j \right)  f(\epsilon_{\bm k})
\label{k integral 1}
\end{align}
with a certain function $f(\epsilon)$. Since directions $i=1,\dots,m$ are equivalent
and the second term in the parentheses vanishes due to a cancellation between $k_i$ and $-k_i$, 
we can simplify Eq.~(\ref{k integral 1}) as
\begin{align}
&=
-\frac{4t^2}{m^2}\sum_{\bm k} m\sin^2 k_x f(\epsilon_{\bm k}).
\end{align}
We can symmetrize the momenta $\{k_i\}$ in the integrand due to the cubic symmetry to have
\begin{align}
&=
-\frac{4t^2}{m^2}\sum_{\bm k} \frac{m}{d}\sum_{i=1}^d \sin^2 k_i
f(\epsilon_{\bm k})
\notag
\\
&=
-\frac{4t^2}{m^2}\sum_{\bm k} \frac{m}{d}\left(\sum_{i=1}^d \sin k_i\right)^2
f(\epsilon_{\bm k}),
\label{k integral 2}
\end{align}
%$\sum_i \sin^2 k_i$ in Eq.~(\ref{k integral 2}) can be replaced by $d/2$ in the limit of $d\to\infty$ 
%since $\sum_i \sin^2 k_i=\sum_i (1-\cos 2k_i)/2=d/2+O(\sqrt{d})$, while
where $4t^2(\sum_{i=1}^d \sin k_i)^2$ can be replaced by $t^\ast{}^2$
using the joint density of states\cite{TurkowskiFreericks2005} $D(\epsilon,\bar\epsilon)=D(\epsilon)D(\bar\epsilon)$ 
(with $\bar\epsilon=2t\sum_i \sin k_i$)
and $\int d\bar\epsilon D(\bar\epsilon)\bar\epsilon^2=t^\ast {}^2$.
Equation (\ref{k integral 1}) is finally reduced to
\begin{align}
\sum_{\bm k} \dot\epsilon_{\bm k} \dddot\epsilon_{\bm k} f(\epsilon_{\bm k}) =
-\frac{t^\ast{}^2}{d^2 \alpha} \sum_{\bm k} f(\epsilon_{\bm k})
\end{align}
in the infinite-dimensional limit. The resulting form can be expressed as a single integral of a function of $\epsilon=\epsilon_{\bm k}$,
which can be evaluated analytically in terms of the local Green's function and self-energy.
Similar simplifications apply to all the possible terms in the THG susceptibility.
In Appendix \ref{momentum integral} we summarize some useful formulae to simplify various types of momentum integrals.

\section{Dotted DMFT}
\label{dotted dmft}

In this section, we propose an efficient way to calculate the vertex correction for nonlinear dynamical susceptibilities,
which we call the ``dotted DMFT'' (Fig.~\ref{dotted DMFT algorithm}). 
This is because the method enables us to evaluate derivatives of Green's function and the self-energy
with respect to the external field, which are required to obtain nonlinear response functions.
We explain the formulation in the context of third-harmonic generation here,
but it can be generalized to arbitrary dynamical susceptibilities.

To start with, let us assume that the system
reaches the (time-periodic) nonequilibrium steady state in the long-time limit
in the presence of an ac electric field. The steady state emerges due to the balance
between continuous excitations by the electric field and an energy dissipation to a heat bath,
i.e., the system considered must be an open system.
%(at least at the starting point).

In the time-periodic nonequilibrium steady state, the time translational symmetry is partially recovered for
Green's function, $G(t+\mathcal T,t'+\mathcal T)=G(t,t')$ (with $\mathcal T=2\pi/\Omega$ being the period of the driving field). 
In principle,
one can determine the interacting Green's function within the nonequilibrium steady-state DMFT, or Floquet DMFT,
\cite{noneqDMFTreview,TsujiOkaAoki2008}
which is capable of treating an arbitrarily large amplitude of the electric field.
For the present purpose, on the other hand, it is sufficient to calculate Green's function up to the third order 
in the driving field.

This motivates us to expand Green's function and the self-energy with respect to
the driving field,
\begin{align}
G(t,t')
&=
G_{\rm eq}(t,t')
+\dot G(t,t';\Omega) A e^{-i\Omega t'}
\notag
\\
&\quad
+\frac{1}{2}\ddot G(t,t';\Omega) A^2 e^{-2i\Omega t'}
+\cdots,
\label{G expansion}
\\
\Sigma(t,t')
&=
\Sigma_{\rm eq}(t,t')
+\dot \Sigma(t,t';\Omega) A e^{-i\Omega t'}
\notag
\\
&\quad
+\frac{1}{2}\ddot \Sigma(t,t';\Omega) A^2 e^{-2i\Omega t'}
+\cdots.
\label{Sigma expansion}
\end{align}
Here $G_{\rm eq}(t,t')$ and $\Sigma_{\rm eq}(t,t')$ are the equilibrium Green's function and self-energy, respectively,
and the external field is assumed to be in a form of $\bm A(t)={\bm e}Ae^{-i\Omega t}$.
If one considers a real field such as $\bm A(t)={\bm e}A\cos \Omega t$, one has to extend the expansion including
cross terms between $e^{-i\Omega t}$ and $e^{i\Omega t}$.
There is an ambiguity in the definition of the expansion coefficients: the factor $e^{-in\Omega t'}$ in the $n$th order
can be replaced by $e^{-in\Omega[x t+(1-x)t']}$ ($x\in\mathbb R$). This is possible as long as the condition
$G(t+\mathcal T,t'+\mathcal T)=G(t,t')$ holds. In this paper, we adopt the convention with $x=0$.

The advantage of expanding Green's function with respect to $A$
rather than directly treating the nonequilibrium Green's function 
is that the full time-translation symmetry is available at each order in the expansion.
To see this, let us write the expansion as
\begin{align}
G(t,t')
&=
G_{\rm eq}(t,t')+G^{(1)}(t,t';\Omega)A+\frac{1}{2}G^{(2)}(t,t';\Omega)A^2+\cdots.
\end{align}
At the $n$th order, the term contains multiple of $n$ ac fields, so that it acquires the phase $e^{-in\Omega \bar t}$
when time translation $t\to t+\bar t$ is operated. For example, for $n=2$ we have
\begin{align}
G^{(2)}(t+\bar t,t'+\bar t;\Omega)
&=
e^{-2i\Omega\bar t} G^{(2)}(t,t';\Omega).
\end{align}
Since $G^{(2)}(t,t')=\ddot G(t,t') e^{-2i\Omega t'}$ by definition,
$\ddot G$ becomes time-translation invariant:
\begin{align}
\ddot G(t+\bar t,t'+\bar t;\Omega)
&=
\ddot G(t,t';\Omega).
\end{align}
The same applies to all orders. This allows us to write $\ddot G(t,t';\Omega)$ as
a single-time function $\ddot G(t-t';\Omega)\equiv \ddot G(t,t';\Omega)$,
which can be Fourier transformed as
\begin{align}
\ddot G(t-t';\Omega)
&=
\int \frac{d\omega}{2\pi} e^{-i\omega(t-t')}\ddot G(\omega;\Omega).
\end{align}
This is a great advantage because it is no longer necessary to treat the two-time Green's function $G(t,t')$
in favor of a single-frequency function. The following formulation can be implemented 
in the same way as in equilibrium which enjoys the full time-translation symmetry.

Now the task is to evaluate the expansion coefficients order by order.
When the system has an inversion symmetry,
the local Green's function and self-energy must be parity even, 
%(the fact that the self-energy 
%is parity-even essentially relies on the DMFT assumption that the self-energy be spatially local),
while the electric field
is parity odd. This implies that odd-order expansion coefficients identically vanish.
The leading contribution then comes from the second order.

In order to determine the expansion coefficients, we differentiate every DMFT self-consistency equation
with respect to $A$, and extract the second-order coefficients. 
Let us start with 
the lattice Dyson equation, $G=\sum_{\bm k} (G_{0\bm k}^{-1}-\Sigma)^{-1}$.
If we take a double derivative with respect to $A$ on both sides of the equation (and use $\dot\Sigma=0$), 
we end up with the ``dotted lattice Dyson equation'',
%\begin{widetext}
\begin{align}
\ddot G^{R,A,<,>}(\omega;\Omega)
=
\sum_{\bm k} \big\{
G_{\bm k}(\omega+2\Omega)[\ddot\epsilon_{\bm k}+\ddot\Sigma(\omega;\Omega)]
G_{\bm k}(\omega)
\notag
\\
\quad
+2G_{\bm k}(\omega+2\Omega)\dot\epsilon_{\bm k} G_{\bm k}(\omega+\Omega)
\dot\epsilon_{\bm k} G_{\bm k}(\omega)
\big\}^{R,A,<,>},
\label{dotted lattice Dyson}
%\\
%\ddot G^{<,>}(\omega;\Omega)
%&=
%\sum_{\bm k} \big\{
%G_{\bm k}^{R}(\omega+2\Omega)[\ddot\epsilon_{\bm k}+\ddot\Sigma^{R}(\omega;\Omega)]
%G_{\bm k}^{<,>}(\omega)
%+G_{\bm k}^{R}(\omega+2\Omega)\ddot\Sigma^{<,>}(\omega;\Omega)
%G_{\bm k}^{A}(\omega)
%+G_{\bm k}^{<,>}(\omega+2\Omega)[\ddot\epsilon_{\bm k}+\ddot\Sigma^{A}(\omega;\Omega)]
%G_{\bm k}^{A}(\omega)
%\big\}
%\notag
%\\
%&\quad
%+2\sum_{\bm k} \big[G_{\bm k}^R(\omega+2\Omega)\dot\epsilon_{\bm k} G_{\bm k}^R(\omega+\Omega)
%\dot\epsilon_{\bm k} G_{\bm k}^{<,>}(\omega)
%+G_{\bm k}^R(\omega+2\Omega)\dot\epsilon_{\bm k} G_{\bm k}^{<,>}(\omega+\Omega)
%\dot\epsilon_{\bm k} G_{\bm k}^{A}(\omega)
%\notag
%\\
%&\quad
%+G_{\bm k}^{<,>}(\omega+2\Omega)\dot\epsilon_{\bm k} G_{\bm k}^{A}(\omega+\Omega)
%\dot\epsilon_{\bm k} G_{\bm k}^{A}(\omega)
%\big],
%\label{dotted lattice Dyson2}
\end{align}
where $R,A,<$, and $>$ denote the retarded, advanced, lesser, and greater components
of nonequilibrium Green's functions, respectively. 
For the detailed definition we refer to Ref.~\onlinecite{noneqDMFTreview}. 
For the notation of $R, A, <$, and $>$ for products of 
nonequilibrium Green's functions, see Appendix \ref{notation}.
Note that
when Green's function has a matrix form (as in the superconducting state),
$\ddot {\hat G}^A(\omega;\Omega)\neq [\ddot {\hat G}^R(\omega;\Omega)]^\dagger$, so that
the advanced component has to be calculated independently of the retarded one.
%The form of the lesser and greater components follow the Langreth rule.\cite{Langreth1976}

Similarly, we differentiate the impurity Dyson equation, $G=(\mathcal G_0^{-1}-\Sigma)^{-1}$ (with $\mathcal G_0$ being the Weiss Green's function) twice with respect to $A$
to obtain the ``dotted impurity Dyson equation'',
\begin{align}
\ddot G^{R,A,<,>}(\omega;\Omega)
&=
-\{G(\omega+2\Omega)[(\ddot{\mathcal G}_0^{-1})(\omega;\Omega)
\notag
\\
&\quad
-\ddot\Sigma(\omega;\Omega)]G(\omega)\}^{R,A,<,>}.
%\\
%\ddot G^{<,>}(\omega;\Omega)
%&=
%-G^{R}(\omega+2\Omega)[(\ddot{\mathcal G}_0^{-1})^R(\omega;\Omega)
%-\ddot\Sigma^{R}(\omega;\Omega)]G^{<,>}(\omega)
%-G^{R}(\omega+2\Omega)[(\ddot{\mathcal G}_0^{-1})^{<,>}(\omega;\Omega)
%-\ddot\Sigma^{<,>}(\omega;\Omega)]G^{A}(\omega)
%\notag
%\\
%&\quad
%-G^{<,>}(\omega+2\Omega)[(\ddot{\mathcal G}_0^{-1})^{A}(\omega;\Omega)
%-\ddot\Sigma^{A}(\omega;\Omega)]G^{A}(\omega).
\end{align}
Here the double-dotted inverse of the Weiss Green's functions reads
\begin{align}
(\ddot{\mathcal G}_0^{-1})^{R,A,<,>}(\omega;\Omega)
&=
-[\mathcal G_0^{-1}(\omega+2\Omega)\ddot{\mathcal G}_0(\omega;\Omega)
\mathcal G_0^{-1}(\omega)]^{R,A,<,>}.
%\\
%(\ddot{\mathcal G}_0^{-1})^{<,>}(\omega;\Omega)
%&=
%-(\ddot{\mathcal G}_0^{-1})^{R}(\omega;\Omega)\mathcal G_0^{<,>}(\omega)
%(\mathcal G_0^{A})^{-1}(\omega)
%-(\mathcal G_0^{-1})^{R}(\omega+2\Omega)\ddot{\mathcal G}_0^{<,>}(\omega;\Omega)
%(\mathcal G_0^{A})^{-1}(\omega)
%\notag
%\\
%&\quad
%-(\mathcal G_0^{-1})^{R}(\omega+2\Omega)\mathcal G_0^{<,>}(\omega+2\Omega)
%(\ddot{\mathcal G}_0^{-1})^{A}(\omega;\Omega).
\end{align}
%\end{widetext}

\begin{figure}[t]
\begin{center}
\includegraphics[width=8cm]{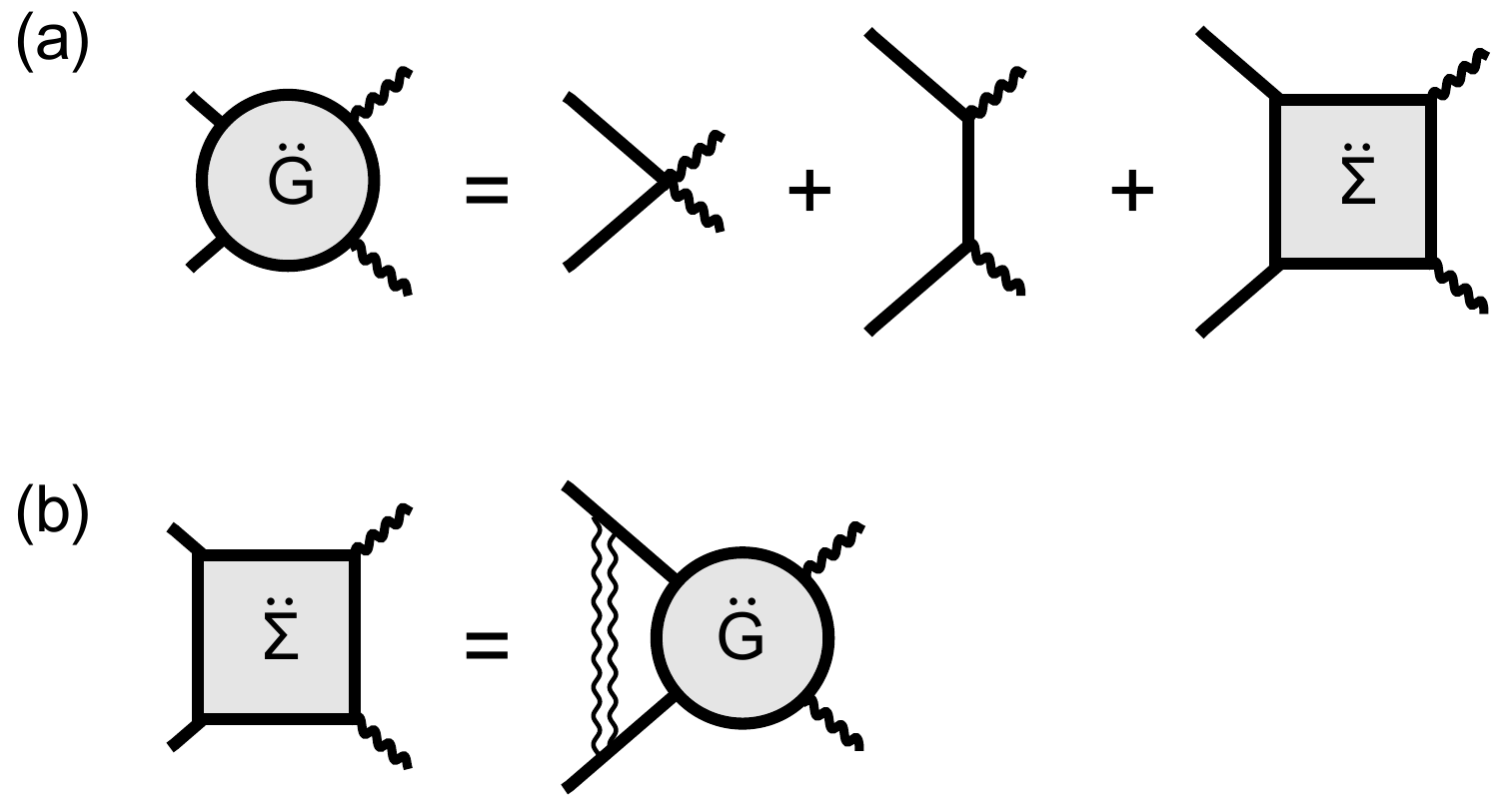}
\caption{
Diagrammatic representations of the dotted lattice Dyson equation (\ref{dotted lattice Dyson}) (a), and the dotted Migdal approximation (\ref{dotted Migdal}) (b).
The solid, wavy, and double-wavy lines represent the electron, photon, and phonon propagators, respectively.}
\label{migdal diagram}
\end{center}
\end{figure}

To close the equation for the dotted functions, we need an explicit (diagrammatic) solution for the nonequilibrium impurity problem.
This depends on the model and approximation. In the present case of the Holstein model, 
we employ the (unrenormalized) Migdal approximation,\cite{Kemper2015,Sentef2015,Murakami2016a,Murakami2016b}
\begin{align}
\Sigma^{<,>}(t,t')
&=
ig^2 D_0^{<,>}(t,t')G^{<,>}(t,t'),
\end{align}
which assumes that phonons stay in equilibrium.
After expanding $\Sigma$ and $G$ with respect to $A$ as in Eqs.~(\ref{G expansion}) and (\ref{Sigma expansion}), 
and compare both sides of the equation at the order $A^2$, we obtain
\begin{align}
\ddot\Sigma^{<,>}(t;\Omega)
&=
ig^2D_0^{<,>}(t)\ddot G^{<,>}(t;\Omega).
\label{dotted Migdal}
\end{align}
The corresponding retarded and advanced components are given by
\begin{align}
\ddot\Sigma^{R}(t;\Omega)
&=
\theta(t)[\ddot\Sigma^>(t;\Omega)-\ddot\Sigma^<(t;\Omega)],
\\
\ddot\Sigma^{A}(t;\Omega)
&=
\theta(-t)[\ddot\Sigma^<(t;\Omega)-\ddot\Sigma^>(t;\Omega)],
\end{align}
where $\theta(t)=1$ for $t\ge 0$ and $=0$ otherwise (step function).
In this way, the dotted DMFT naturally generates the impurity solver
that is consistent with the approximation used in the equilibrium DMFT.
The diagrammatic representation of the dotted lattice Dyson equation (\ref{dotted lattice Dyson})
and dotted Migdal approximation (\ref{dotted Migdal}) are shown in Fig.~\ref{migdal diagram}.
In particular, if we expand the dotted self-energy only with $G$ and $D_0$,
it has a ladder structure, which represents the contribution from the amplitude mode.\cite{Murakami2016a}

The technique can be applied to any models in principle, as far as the diagrammatic expression for the impurity problem is given.
To see how it works further, let us take another prototypical example, the Hubbard model,
\begin{align}
H&=
\sum_{ij} t_{ij} c_{i\sigma}^\dagger c_{j\sigma}
+U\sum_i n_{i\uparrow} n_{i\downarrow},
\end{align}
where $U$ is the on-site Coulomb interaction. 
The Hubbard model is of particular interest in its own right from the point of view of
large nonlinear optical responses of Mott insulators.\cite{Kishida2000,Kishida2001}
The diagrammatic approximation often used
for the nonequilibrium impurity problem is the iterative perturbation theory (IPT),\cite{GeorgesKotliar1992,GeorgesKotliarKrauthRozenberg1996}
\begin{align}
\Sigma_{\sigma}^{<,>}(t,t')
&=
U^2 \mathcal G_{0,\sigma}^{<,>}(t,t') \mathcal G_{0,-\sigma}^{>,<}(t',t) \mathcal G_{0,-\sigma}^{<,>}(t,t'),
\end{align}
which is nothing but the bare second-order weak-coupling perturbation theory.
The dotted impurity solution derived from IPT is given as
\begin{align}
\ddot\Sigma_{\sigma}^{<,>}(t;\Omega)
&=
U^2 \ddot{\mathcal G}_{0,\sigma}^{<,>}(t;\Omega) \mathcal G_{0,-\sigma}^{>,<}(-t) \mathcal G_{0,-\sigma}^{<,>}(t)
\notag
\\
&\quad
+U^2 e^{-2i\Omega t} \mathcal G_{0,\sigma}^{<,>}(t) \ddot{\mathcal G}_{0,-\sigma}^{>,<}(-t;\Omega) \mathcal G_{0,-\sigma}^{<,>}(t)
\notag
\\
&\quad
+U^2 \mathcal G_{0,\sigma}^{<,>}(t) \mathcal G_{0,-\sigma}^{>,<}(-t) \ddot{\mathcal G}_{0,-\sigma}^{<,>}(t;\Omega).
\label{dotted IPT}
\end{align}
Note that the second term in Eq.~(\ref{dotted IPT}) acquires a phase factor $e^{-2i\Omega t}$, since the definition of
the dotted function [Eqs.~(\ref{G expansion}), (\ref{Sigma expansion})] is asymmetric between $t$ and $t'$.
In the symmetric case (i.e., $x=1/2$), the phase factor does not appear.

Combining the dotted impurity solution with the dotted lattice and impurity Dyson equations, we can determine
$\ddot G$ and $\ddot\Sigma$ self-consistently.
We summarize the algorithm flow for the dotted DMFT in Fig.~\ref{dotted DMFT algorithm}.
First, we solve the equilibrium DMFT in the real-time (real-frequency) formalism 
(the upper part of Fig.~\ref{dotted DMFT algorithm}).
Once the self-consistency loop is converged, we move on to the second step of calculating the dotted functions
(the lower part of Fig.~\ref{dotted DMFT algorithm}).
We fix the external frequency $\Omega$, and iteratively solve the dotted DMFT self-consistency loop.
This process runs for every chosen $\Omega$. Thus the computational cost for the dotted DMFT is roughly
$N_\Omega$ times that for the equilibrium DMFT, where $N_\Omega$ is the number of $\Omega$ values taken.

\begin{figure}[t]
\begin{center}
\includegraphics[width=8cm]{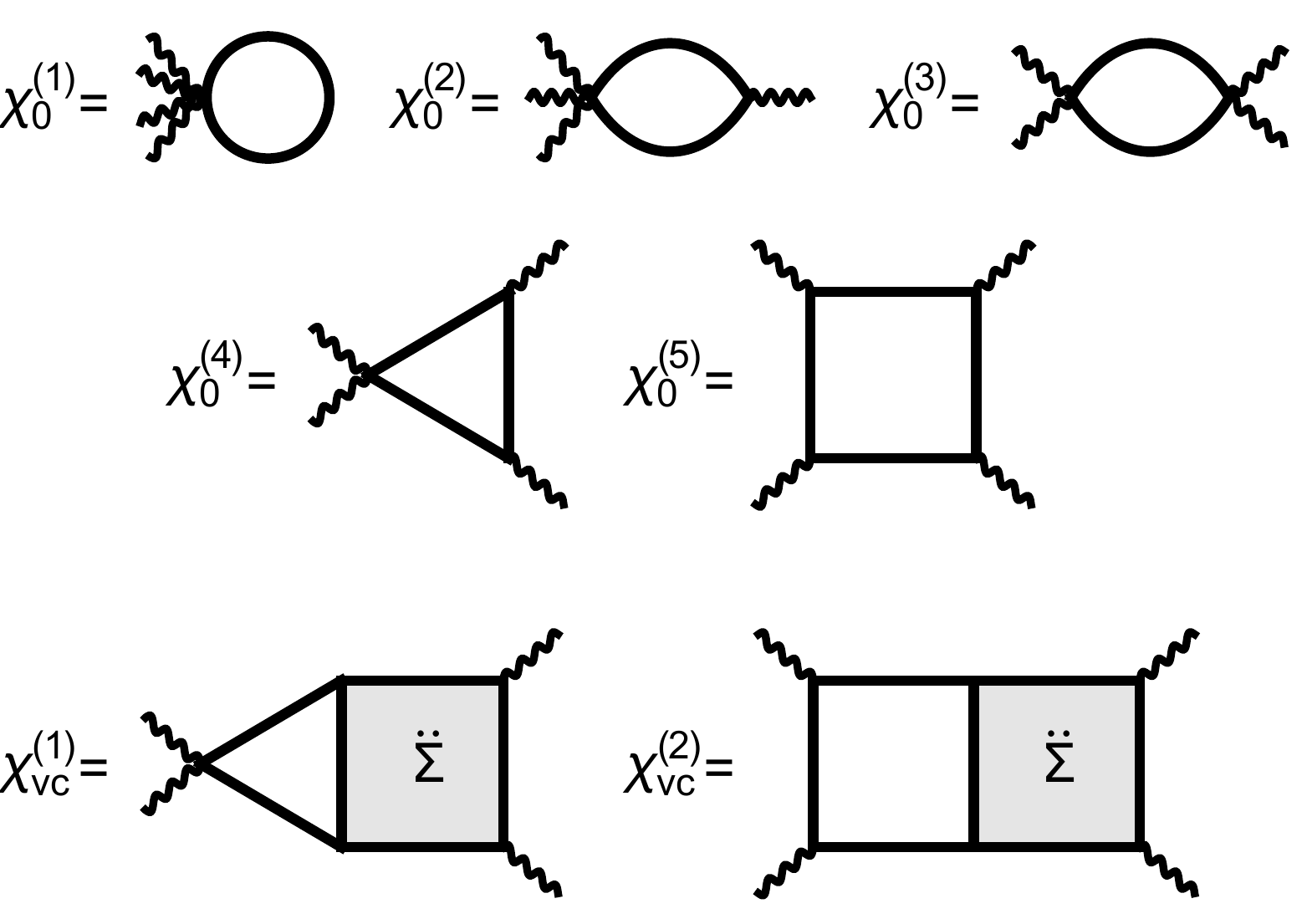}
\caption{Feynman diagrams for the susceptibility for the third-harmonic generation. 
There are five (two) topologically different diagrams
for the bare susceptibility $\chi_0$ (vertex correction $\chi_{\rm vc}$).
Solid and wavy lines represent the electron and external photon propagators, respectively,
while the shaded box represents the vertex correction $\ddot\Sigma$.
Among the four photon lines,
one is outgoing with an energy $3\Omega$, while the other three are incoming with an energy $\Omega$.
The photon lines attached directly to the vertex are incoming.}
\label{THG diagram}
\end{center}
\end{figure}

After collecting the results for the set of $\Omega$s, we can calculate the THG susceptibility (\ref{THG susceptibility}),
which is obtained from the third derivative of the current (\ref{current}) with respect to $A$.
%for the third-harmonic generation (THG).
%It is defined as the third-order current response to the external vector potential $\bm A(t)=\hat{\bm e} A e^{-i\Omega t}$
%($\hat{\bm e}$ is the unit polarization vector),
%\begin{align}
%j^{(3)}(t)
%&=
%\chi(\Omega)A^3 e^{-3i\Omega t}.
%\end{align}
%Since the vector potential and electric field are related via $\bm E(t)=-\frac{\partial\bm A(t)}{\partial t}$,
%we have $A=E/(i\Omega)$. Thus the physical THG intensity observed in experiments is given by
%\begin{align}
%I&=
%\frac{|\chi(\Omega)|^2}{\Omega^6}.
%\end{align}
The THG susceptibility is classified into the bare susceptibility $\chi_0$ and vertex correction $\chi_{\rm vc}$,
\begin{align}
\chi_{\rm THG}(\Omega)
&=
\chi_0(\Omega)+\chi_{\rm vc}(\Omega),
\end{align}
according as whether it contains the derivative of the self-energy ($\ddot\Sigma$).
Each term is further decomposed respectively as
$\chi_0(\Omega)=\sum_{i=1}^5 \chi_0^{(i)}(\Omega)$ and
$\chi_{\rm vc}(\Omega)=\sum_{i=1}^2 \chi_{\rm vc}^{(i)}(\Omega)$,
following the topological classification of the corresponding Feynman diagrams as displayed in Fig.~\ref{THG diagram}.
Combining Figs.~\ref{migdal diagram} and \ref{THG diagram}, one can see that
$\chi_{\rm vc}^{(1)}$ contains the non-resonant [Fig.~\ref{raman diagram}(a)] and mixed [Fig.~\ref{raman diagram}(b)] diagrams,
while $\chi_{\rm vc}^{(2)}$ contains the mixed [Fig.~\ref{raman diagram}(b)] and resonant [Fig.~\ref{raman diagram}(c)] ones.
In the BCS approximation, only $\chi_0^{(1)}, \chi_0^{(3)}$, and $\chi_{\rm vc}^{(1)}$,
which have the non-resonant coupling to the light,
are nonzero as explained in the introduction, and the rest vanish exactly.
On the other hand, when one goes beyond the BCS approximation all the terms are generally non-vanishing and cannot be neglected,
so that one has to evaluate all of them.
%We can calculate each term by taking the third derivative of the current,
%\begin{align}
%\bm j(t)
%&=
%-i\sum_{\bm k} {\bm v}_{\bm k+\bm A(t)} G_{\bm k}^<(t,t),
%\end{align}
%with respect to $A$. We measure the current along the direction of $\hat{\bm e}$
%parallel to the electric field,
%\begin{align}
%j(t)&=\bm j(t)\cdot \hat{\bm e}.
%\end{align}

The explicit form of the bare susceptibilities are the following:
\begin{widetext}
\begin{align}
\chi_0^{(1)}(\Omega)
&=
-\frac{i}{6}\sum_{\bm k} \int \frac{d\omega}{2\pi}
[\ddddot\epsilon_{\bm k} G_{\bm k}(\omega)]^<,
\label{chi0 1}
\\
\chi_0^{(2)}(\Omega)
&=
-\frac{i}{2}\sum_{\bm k} \int \frac{d\omega}{2\pi}
[\dddot\epsilon_{\bm k}G_{\bm k}(\omega+\Omega)
\dot\epsilon_{\bm k}G_{\bm k}(\omega)]^<
-\frac{i}{6}\sum_{\bm k} \int \frac{d\omega}{2\pi}
[\dot\epsilon_{\bm k}G_{\bm k}(\omega+3\Omega)
\dddot\epsilon_{\bm k}G_{\bm k}(\omega)]^<,
\label{chi0 2}
\\
\chi_0^{(3)}(\Omega)
&=
-\frac{i}{2}\sum_{\bm k} \int \frac{d\omega}{2\pi}
[\ddot\epsilon_{\bm k}G_{\bm k}(\omega+2\Omega)
\ddot\epsilon_{\bm k}G_{\bm k}(\omega)]^<,
\label{chi0 3}
\\
\chi_0^{(4)}(\Omega)
&=
-i\sum_{\bm k} \int \frac{d\omega}{2\pi}
[\ddot\epsilon_{\bm k}G_{\bm k}(\omega+2\Omega)
\dot\epsilon_{\bm k}G_{\bm k}(\omega+\Omega)
\dot\epsilon_{\bm k}G_{\bm k}(\omega)]^<
\notag
\\
&\quad
-\frac{i}{2}\sum_{\bm k} \int \frac{d\omega}{2\pi}
[\dot\epsilon_{\bm k}G_{\bm k}(\omega+3\Omega)
\ddot\epsilon_{\bm k}G_{\bm k}(\omega+\Omega)
\dot\epsilon_{\bm k}G_{\bm k}(\omega)
+\dot\epsilon_{\bm k}G_{\bm k}(\omega+3\Omega)
\dot\epsilon_{\bm k}G_{\bm k}(\omega+2\Omega)
\ddot\epsilon_{\bm k}G_{\bm k}(\omega)]^<,
\label{chi0 4}
\\
\chi_0^{(5)}(\Omega)
&=
-i\sum_{\bm k} \int \frac{d\omega}{2\pi}
[\dot\epsilon_{\bm k}G_{\bm k}(\omega+3\Omega)
\dot\epsilon_{\bm k}G_{\bm k}(\omega+2\Omega)
\dot\epsilon_{\bm k}G_{\bm k}(\omega+\Omega)
\dot\epsilon_{\bm k}G_{\bm k}(\omega)]^<.
\label{chi0 5}
\end{align}
For the notation of $<$ for products of 
nonequilibrium Green's functions, see Appendix \ref{notation}.
Note that $\chi_0^{(i)}$ ($i=1,\dots,5$) do not contain $\ddot\Sigma$, so that they can be computed independently of the dotted DMFT.
The vertex corrections are also explicitly derived as
\begin{align}
\chi_{\rm vc}^{(1)}(\Omega)
&=
-\frac{i}{2}\sum_{\bm k} \int \frac{d\omega}{2\pi}
[\ddot\epsilon_{\bm k} G_{\bm k}(\omega+2\Omega)\ddot\Sigma(\omega;\Omega)G_{\bm k}(\omega)]^<,
\label{chi vc1}
\\
\chi_{\rm vc}^{(2)}(\Omega)
&=
-\frac{i}{2}\sum_{\bm k} \int \frac{d\omega}{2\pi}
[\dot\epsilon_{\bm k} G_{\bm k}(\omega+3\Omega)\ddot\Sigma(\omega+\Omega;\Omega)
G_{\bm k}(\omega+\Omega) \dot\epsilon_{\bm k} G_{\bm k}(\omega)
+\dot\epsilon_{\bm k} G_{\bm k}(\omega+3\Omega)\dot\epsilon_{\bm k} 
G_{\bm k}(\omega+2\Omega) \ddot\Sigma(\omega;\Omega) G_{\bm k}(\omega)]^<.
\label{chi vc2}
\end{align}
\end{widetext}
Using the momentum integral formulae listed in Appendix \ref{momentum integral}, 
one can show that $\chi_0^{(4)}$, $\chi_0^{(5)}$, $\chi_{\rm vc}^{(1)}$, and $\chi_{\rm vc}^{(2)}$
do not depend on the polarization parameter $\alpha$, while $\chi_0^{(1)}$, $\chi_0^{(2)}$, and $\chi_0^{(3)}$ do.

The algorithm can be generalized to arbitrary higher-order derivatives;
they are determined in turn from lower orders
[$(G,\Sigma) \to (\ddot G, \ddot\Sigma) \to (\ddddot G, \ddddot\Sigma)$]
due to the hierarchical structure of the dotted DMFT self-consistency
defined at each derivative order. 
Essentially the same method has been used to evaluate the optical conductivity
for periodically driven systems in Floquet DMFT,\cite{TsujiOkaAoki2009} where $\dot G$ and $\dot\Sigma$ are needed.
The first derivative can be nonzero, since the parity symmetry is broken by the presence of the external driving field.

So far we have formulated the dotted DMFT for the normal phase, but
it is straightforward to extend the approach to the superconducting phase. There, we have to impose the following modifications:
the electron Green's functions and self-energy should be represented by $2\times 2$ matrices (Nambu-Gor'kov formalism),
$\epsilon_{\bm k}, \ddot\epsilon_{\bm k}$, and $\ddddot\epsilon_{\bm k}$ appearing in the dotted DMFT should be multiplied by
$\tau_3$ (the third component of the Pauli matrix),
and the dotted impurity solution (\ref{dotted Migdal}) for the Holstein model should be replaced by
\begin{align}
\ddot{\hat\Sigma}^{<,>}(t;\Omega)
&=
ig^2D_0^{<,>}(t)\tau_3\ddot{\hat G}^{<,>}(t;\Omega)\tau_3.
\end{align}

Let us finally comment on the generality of the present formulation. Although we describe the dotted DMFT formulation
for the THG susceptibility, it is not restricted to THG but can be generalized to arbitrary dynamical response functions.
One can introduce an infinitesimal external field (not necessarily an electric field), and take the derivative with respect to it for observables. It may contain a derivative of the self-energy, which can be evaluated by the corresponding dotted DMFT,
where the DMFT self-consistency equations are differentiated with respect to the external field.
For example, the dynamical pair susceptibility,\cite{Cea2014,Murakami2016a}
\begin{align}
\chi_{\rm pair}^R(\Omega)
&=
-i\int_0^\infty dt e^{-i\Omega t} \langle [B_{\bm 0}(t), B_{\bm 0}(0)]\rangle,
\label{chi pair}
\end{align}
is defined as the response of the pairing amplitude $\langle B_{\bm 0}\rangle$
against an external pair potential $H_{\rm ex}(t)=\varepsilon B_{\bm 0}e^{-i\Omega t}$,
where 
\begin{align}
B_{\bm 0}=\sum_i (c_{i\uparrow}^\dagger c_{i\downarrow}^\dagger+c_{i\downarrow}c_{i\uparrow})
\end{align}
is the bosonic pairing operator with the center-of-mass momentum $\bm q=\bm 0$.
This quantity detects collective amplitude oscillations of the superconducting order parameter.
The dotted DMFT is then constructed by differentiating the DMFT self-consistency with respect to
the pair field potential. The resulting dotted lattice Dyson equation reads
\begin{align}
\dot{\hat G}^{R,A,<,>}(\omega;\Omega)
&=
\sum_{\bm k} 
\{\hat G_{\bm k}(\omega+\Omega)[\tau_1+\dot{\hat\Sigma}(\omega;\Omega)]\hat G_{\bm k}(\omega)\}^{R,A,<,>},
%\\
%\dot {\hat G}^{<,>}(\omega;\Omega)
%&=
%\sum_{\bm k} \big\{\hat G_{\bm k}^{R}(\omega+\Omega)[\tau_1+\dot{\hat\Sigma}^{R}(\omega;\Omega)]\hat G_{\bm k}^{<,>}(\omega)
%\notag
%\\
%&\quad
%+\hat G_{\bm k}^{R}(\omega+\Omega)\dot{\hat\Sigma}^{<,>}(\omega;\Omega)\hat G_{\bm k}^{A}(\omega)
%\notag
%\\
%&\quad
%+\hat G_{\bm k}^{<,>}(\omega+\Omega)[\tau_1+\dot{\hat\Sigma}^{A}(\omega;\Omega)]\hat G_{\bm k}^{A}(\omega)
%\big\}.
\end{align}
where we have adopted an extended notation of $(\tau_1)^{R,A}=\tau_1$ and $(\tau_1)^{<,>}=0$.
Once the dotted DMFT is solved, the dynamical pair susceptibility can be calculated as
\begin{align}
\chi_{\rm pair}^R(\Omega)
&=
-i\int \frac{d\omega}{2\pi} {\rm Tr} [\tau_1 \dot {\hat G}^<(\omega;\Omega)].
%-i\sum_{\bm k}\int \frac{d\omega}{2\pi}
%{\rm Tr}[\tau_1 G_{\bm k}^R(\omega+\Omega) \tau_1 G_{\bm k}^<(\omega)
%+\tau_1 G_{\bm k}^<(\omega+\Omega) \tau_1 G_{\bm k}^A(\omega)
%\notag
%\\
%&\quad
%+\tau_1 G_{\bm k}^R(\omega+\Omega) \dot\Sigma^R(\omega;\Omega) G_{\bm k}^<(\omega)
%+\tau_1 G_{\bm k}^R(\omega+\Omega) \dot\Sigma^<(\omega;\Omega) G_{\bm k}^A(\omega)
%+\tau_1 G_{\bm k}^<(\omega+\Omega) \dot\Sigma^A(\omega;\Omega) G_{\bm k}^A(\omega)]
\end{align}
In the next section, we demonstrate the results obtained with the dotted DMFT
for the THG susceptibility along with the dynamical pair susceptibility.

\section{Results}
\label{results}

Let us now turn to the results of the dotted DMFT for the superconducting phase of the Holstein model.
The parameters are taken to be $g=0.8$, $\omega_0=0.6$, 
$\gamma=0.2$, and $\delta=0.005$. This corresponds to the effective interaction of $\lambda=0.77$ [Eq.~(\ref{lambda})],
which is in the moderately correlated regime. The temperature is set to be $T=0.02$, which is low enough
for the system to be in the superconducting state.
The polarization (\ref{polarization}) is set to
a general direction $\alpha=0.5$ without having a bias on the pair breaking effect.

\begin{figure}[t]
\begin{center}
\includegraphics[width=7.6cm]{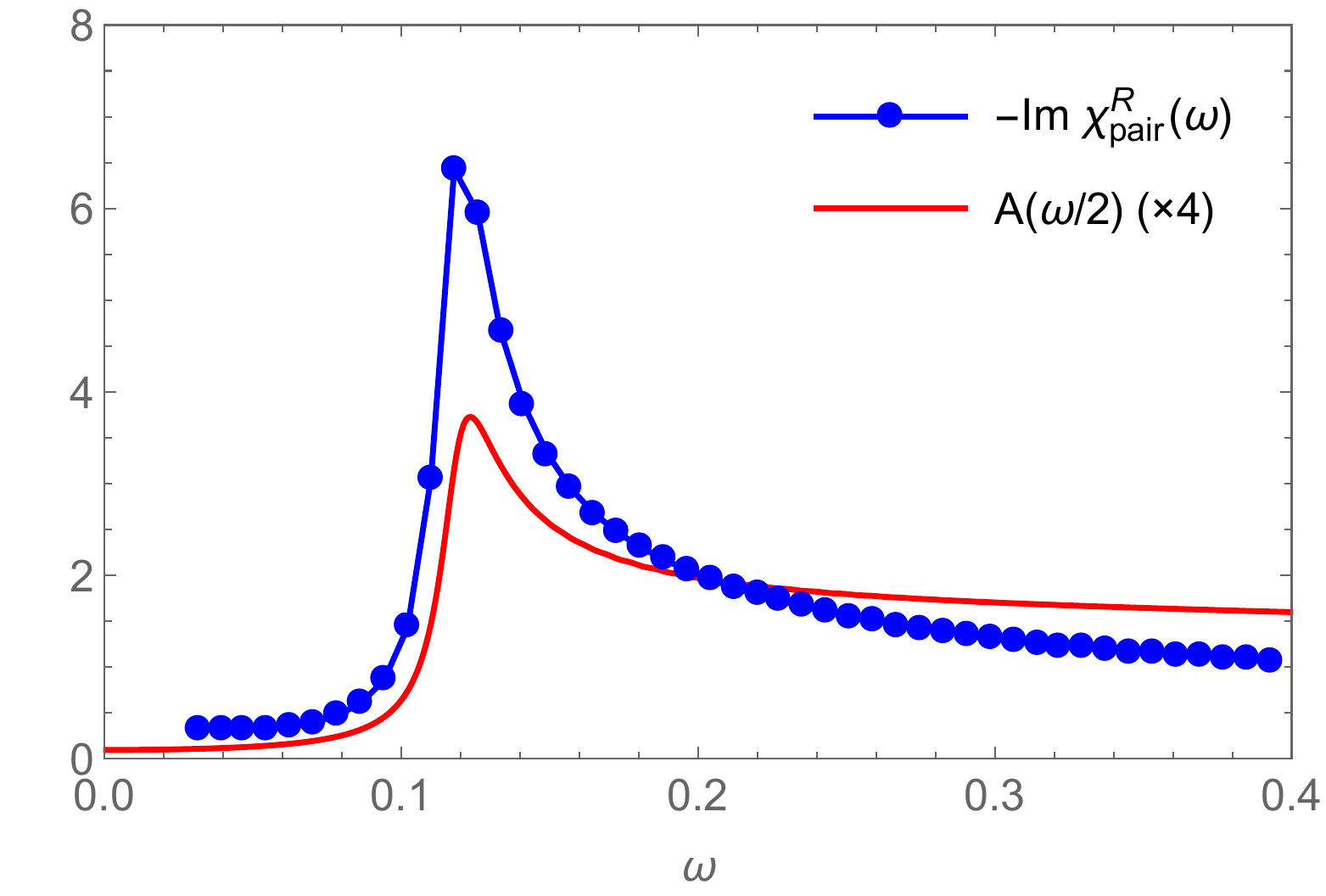}
\caption{The single-particle spectrum $A(\omega/2)$ and 
the dynamical pair susceptibility $-{\rm Im}\,\chi_{\rm pair}^R(\omega)$
calculated by the (dotted) DMFT
for the superconducting phase of the Holstein model with $g=0.8, \omega_0=0.6, T=0.02, \gamma=0.2$, and $\delta=0.005$.}
\label{pair susceptibility}
\end{center}
\end{figure}

In Fig.~\ref{pair susceptibility}, we show the single-particle spectrum $A(\omega)=-{\rm Im}\,G_{11}^R(\omega)/\pi$
(red curve)
along with the dynamical pair susceptibility $-{\rm Im}\, \chi_{\rm pair}^R(\omega)$ (\ref{chi pair}) (blue with the dots),
with the latter calculated by the dotted DMFT. Previously, the dynamical pair susceptibility has been evaluated
from the real-time simulation of the nonequilibrium DMFT,\cite{Murakami2016a} which is one way to avoid
solving the complicated Bethe-Salpeter equation for the vertex correction. 
Here the dotted DMFT serves as an alternative efficient method.

As one can see in Fig.~\ref{pair susceptibility}, the single-particle spectrum shows the superconducting gap
$2\Delta\approx 0.12$ [note that we plot $A(\omega/2)$ in Fig.~\ref{pair susceptibility}]
with the coherence peak at the edge of the band gap. The pair susceptibility also exhibits
a clear gap structure with a resonance peak at $\omega=2\Delta$. 
The result is in agreement with the one previously reported.\cite{Murakami2016a,Murakami2016b}
The resonance peak is produced by the vertex correction, which is immediately confirmed by the comparison
to the bare susceptibility. This suggests that the peak in $\chi_{\rm pair}^R(\omega)$ represents the collective
oscillation of the pairing amplitude with the frequency $2\Delta$, 
which can be identified as the Higgs amplitude mode. The coincidence of the single-particle and two-particle gaps
(up to the factor of 2) holds beyond the BCS approximation, as observed in the previous study.\cite{Murakami2016a,Murakami2016b}

\begin{figure}[t]
\begin{center}
\includegraphics[width=8cm]{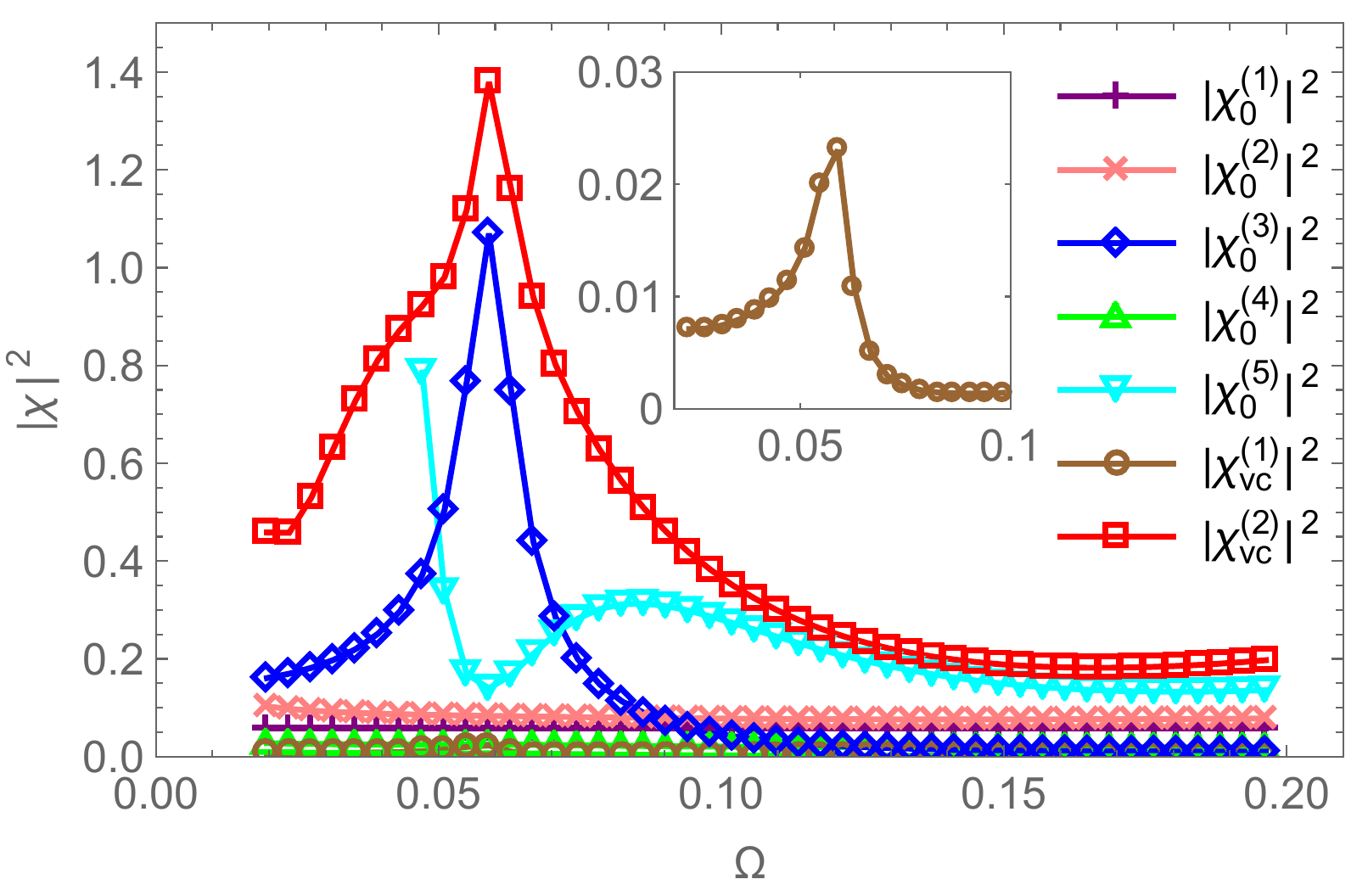}
\caption{The THG susceptibility decomposed into the bare susceptibilities $\chi_0^{(i)}$ ($i=1,\dots,5$)
and vertex corrections $\chi_{\rm vc}^{(i)}$ ($i=1,2$) for the superconducting phase of the Holstein model
calculated with the dotted DMFT.
The parameters are the same as in Fig.~\ref{pair susceptibility}.
The polarization direction of the laser field is taken to be $\alpha=0.5$.
The susceptibilities are normalized by $\chi^\ast=t^\ast/d^2$.
The inset is a blowup of $\chi_{\rm vc}^{(1)}$.}
\label{THG resolved}
\end{center}
\end{figure}

The results of the THG susceptibility $|\chi|^2$ (proportional to the THG intensity observed in experiments)
calculated by the dotted DMFT are plotted in Fig.~\ref{THG resolved}.
%(with the unit of $\chi^\ast=t^\ast/d^2$).
We show the result for each term $\chi_0^{(i)}$ ($i=1,\dots,5$) and $\chi_{\rm vc}^{(i)}$ ($i=1,2$).
First of all, we can see that all the terms contribute to the THG response, which is in sharp contrast
to the BCS approximation where $\chi_0^{(i)}$ ($i=2,4,5$) and $\chi_{\rm vc}^{(2)}$ identically vanish.
In particular, $\chi_0^{(3)}$, $\chi_0^{(5)}$ and $\chi_{\rm vc}^{(2)}$ exhibit dominant contributions.
The resonance peak exists at $\Omega=\Delta\approx 0.06$ in the spectra of $\chi_0^{(3)}$, $\chi_{\rm vc}^{(1)}$, and $\chi_{\rm vc}^{(2)}$. The peak in $\chi_0^{(3)}$ can be interpreted as individual excitations due to Cooper pair breaking,
while the peaks in $\chi_{\rm vc}^{(1)}$ and $\chi_{\rm vc}^{(2)}$ can be interpreted as collective excitations 
resonating with the Higgs amplitude mode, since that is the only known collective mode at energy $2\Delta$.
As expected from the BCS approximation,\cite{Cea2016} the effect of $\chi_{\rm vc}^{(1)}$ is a few orders of magnitude smaller than
that of $\chi_0^{(3)}$ (see the inset of Fig.~\ref{THG resolved}) if one chooses a general polarization direction (here $\alpha=0.5$).
On the other hand, the contribution of $\chi_{\rm vc}^{(2)}$, which has been absent in the BCS approximation,
is quite significant, and can be even larger than that of $\chi_0^{(3)}$. This result implies that 
if one resumes the factors that are not taken into account
in the BCS approximation, such as the retarded nature of the pairing interaction through the electron-phonon coupling,
the Higgs mode can become a prominent component in the THG spectrum. The corrections from the BCS theory are
not necessarily small but can be drastic (at least when the electron-phonon coupling is large enough).
Let us again recall that NbN, which is experimentally used in Refs.~\onlinecite{Matsunaga2013,Matsunaga2014},
has the strong electron-phonon coupling,\cite{Kihlstrom1985,Brorson1990,Chockalingam2008}
so that such corrections from the BCS analysis should be
seriously taken into account.
$\chi_0^{(5)}$ is also not negligible, but this component does not show a resonance with the Higgs mode at $\Omega=\Delta$.
The increase of the spectral weight towards low frequencies (especially for $\chi_0^{(5)}$ and $\chi_{\rm vc}^{(2)}$)
is due to the presence of nonzero $\delta$, with which the system accommodates low-energy excitations.
It can be suppressed when $\delta$ is reduced, so that we can ignore the low-energy features,
although we cannot take the limit of $\delta\to +0$ for the dotted DMFT to be numerically stable.
%since the dotted DMFT becomes numerically unstable when the value of $\delta$ becomes too small.

\begin{figure}[t]
\begin{center}
\includegraphics[width=8cm]{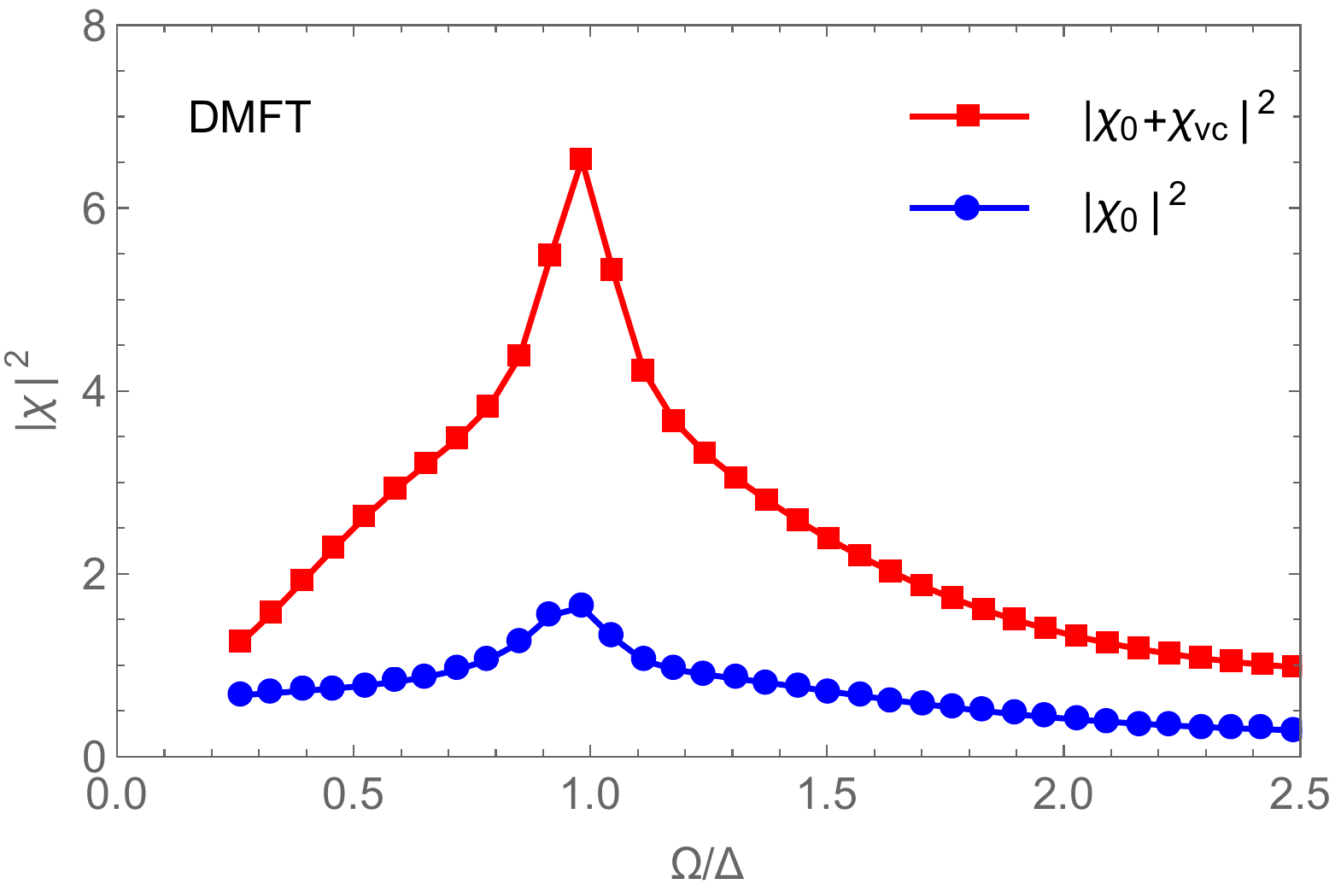}
\caption{The intensity of the third-harmonic generation for the superconducting phase of the Holstein model
calculated by the dotted DMFT. 
The bubble contribution ($\chi_0$) and the total susceptibility including the vertex corrections ($\chi_0+\chi_{\rm vc}$) are plotted.
The parameters are taken to be the same as those of Fig.~\ref{pair susceptibility}.
The polarization direction of the laser field is taken to be $\alpha=0.5$.}
\label{THG}
\end{center}
\end{figure}

Figure~\ref{THG} plots the total THG susceptibility $|\chi|^2=|\chi_0+\chi_{\rm vc}|^2$ as compared with the total bare susceptibility
$|\chi_0|^2$. Here we subtract the low-energy increase
of the spectral weight of $\chi_0^{(5)}$ at $\Omega<0.055$ from $\chi_0$, 
which is out of our interest and could be removed by reducing $\delta$.
We can see that both $\chi$ and $\chi_0$ exhibit conspicuous resonance peaks at $\Omega=\Delta$.
Although the position and shape of the peak do not differ so much between $\chi_0$ and $\chi$, the peak height does.
With the parameters taken here, the height for $\chi$ is enhanced about four times that for $\chi_0$
due to the resonance with the Higgs mode. The main contribution comes from $\chi_{\rm vc}^{(2)}$,
as can be seen in Fig.~\ref{THG resolved}. The resonance width for $\chi_0$ is broadened as compared to that for $\chi_0^{(3)}$ due to the spectral weight of $\chi_0^{(5)}$ distributed around the peak.
The amplitude ratio between $\chi_0$ and $\chi_{\rm vc}$ can depend on
various model parameters (especially we will discuss the phonon-frequency dependence below), 
but at least there is such a possibility in a certain realistic parameter regime
that the vertex correction has a non-negligible effect.

\begin{figure}[t]
\begin{center}
\includegraphics[width=8cm]{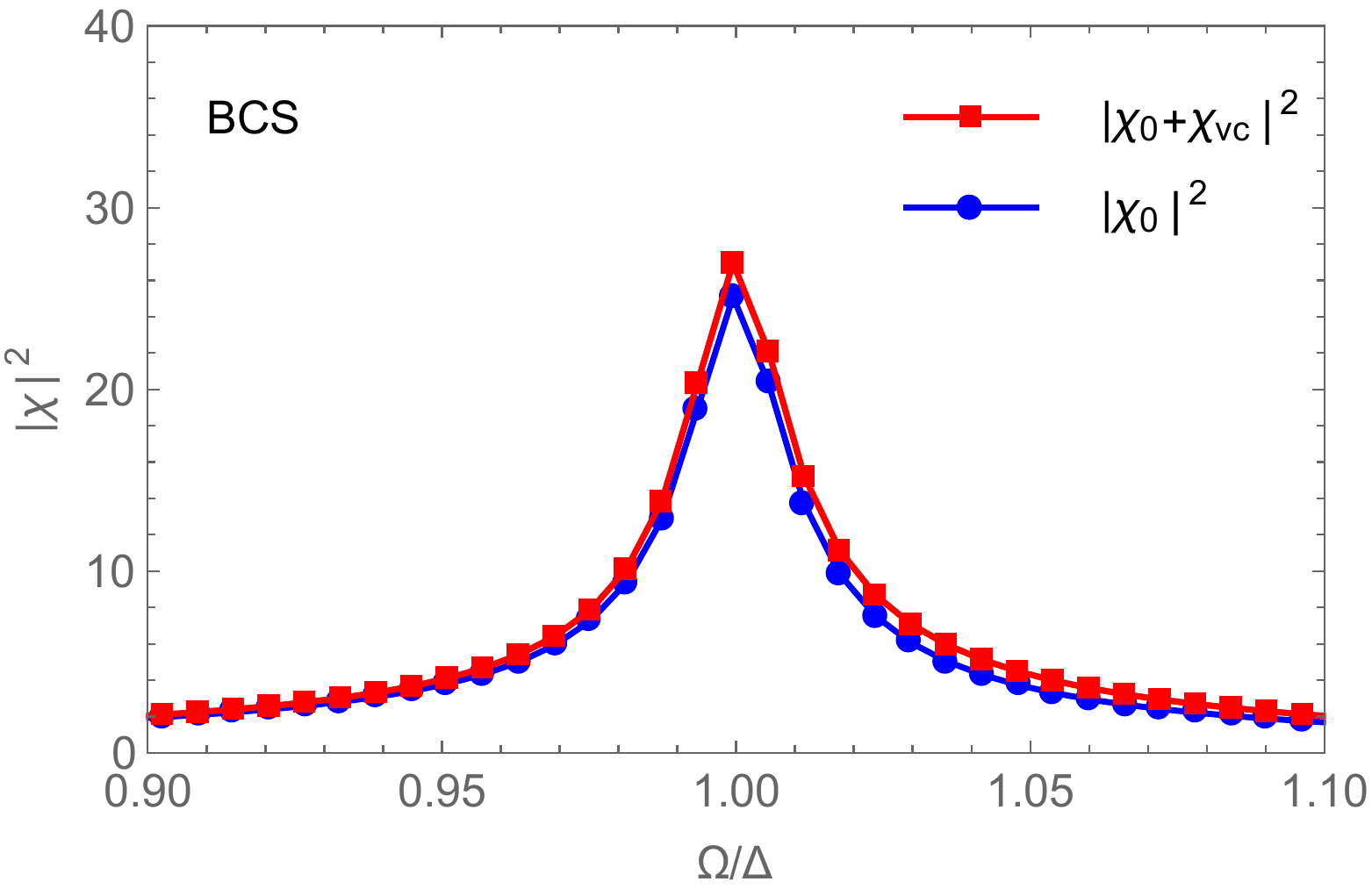}
\caption{The intensity of the third-harmonic generation for the superconducting phase of the Holstein model
calculated within the BCS approximation. 
The bubble contribution ($\chi_0$) and the total susceptibility including the vertex corrections ($\chi_0+\chi_{\rm vc}$) are plotted respectively.
The parameters are taken to be the same as in Fig.~\ref{pair susceptibility}.
The polarization direction of the laser field is taken to be $\alpha=0.5$.
Note the difference in the scales of the axes from Fig.~\ref{THG}.}
\label{THG BCS}
\end{center}
\end{figure}

We are now in position to compare the BCS and DMFT results by 
calculating the THG susceptibility
within the BCS approximation for the same parameter set as those for DMFT. The electron-phonon coupling
is translated into a static attractive interaction via $U=2g^2\omega_0/(\omega_0^2+\gamma^2)$ [see Eqs.~(\ref{U(w)}) and (\ref{D0R})].
In the gap equation, we perform the momentum integral in the range of $|\epsilon_{\bm k}|\le \omega_0$.
The result is displayed in Fig.~\ref{THG BCS}, which indicates that the effect of the vertex correction in BCS is rather small 
for a general polarization direction ($\alpha=0.5$ here). 
The resonance width is much sharper and the peak height is higher in BCS than in DMFT, since the THG susceptibility
diverges at $\Omega=\Delta$ in the limit of $\delta\to 0$ in the BCS theory.
While these are consistent
with the previous studies,\cite{Tsuji2015,Cea2016} the BCS result is markedly different from the DMFT result (Fig.~\ref{THG}) that takes account of
dynamical correlation effects. This is simply because $\chi_{\rm vc}^{(2)}$ is absent in the BCS approximation,
whereas it is generally non-negligible if one considers the retardation in the phonon-mediated interaction (or other effects
that are not included in the BCS approximation such as impurity scattering, Coulomb interaction, etc.).

\begin{figure}[t]
\begin{center}
\includegraphics[width=8cm]{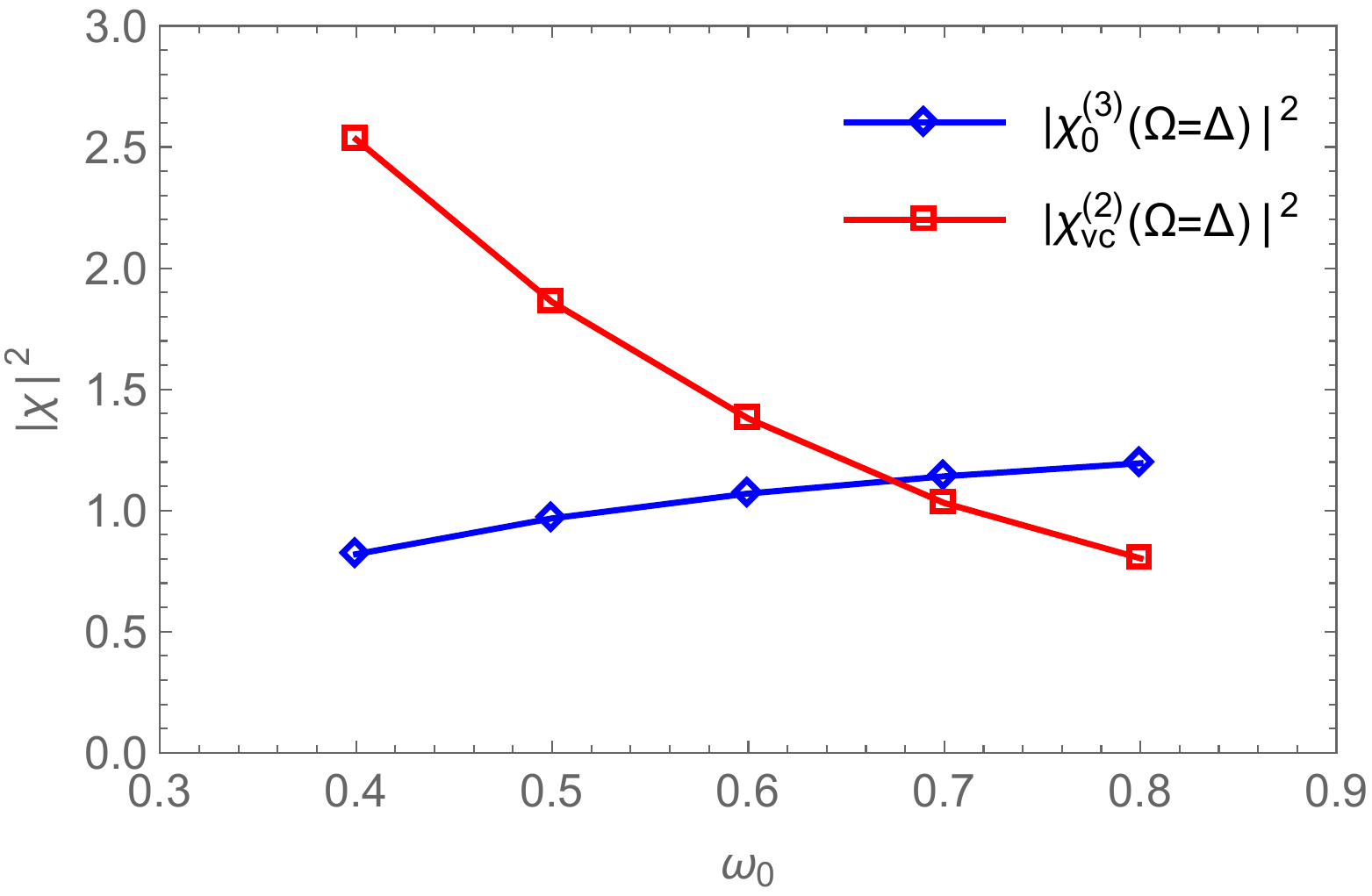}
\caption{Phonon-frequency dependence of $|\chi_0^{(3)}|^2$ and $|\chi_{\rm vc}^{(2)}|^2$ at the resonance ($\Omega=\Delta$)
for the superconducting phase of the Holstein model with $T=0.02$, $\gamma=0.2$, and $\delta=0.005$.
The polarization direction of the laser field is taken to be $\alpha=0.5$.
We tune $g$ for each $\omega_0$ such that the superconducting gap is fixed to a constant ($2\Delta\approx 0.12$).}
\label{w0 dependence}
\end{center}
\end{figure}

To confirm that the retardation effect is essential in enhancing the contribution of the Higgs mode to
the THG resonance, we calculate $\omega_0$ dependence of the THG susceptibility. Here we focus on
$\chi_0^{(3)}(\Omega)$ and $\chi_{\rm vc}^{(2)}(\Omega)$ that are in charge of the resonance structures at $\Omega=\Delta$.
A systematic comparison between the susceptibilities at different $\omega_0$ is made by tuning 
the electron-phonon coupling $g$ such that the superconducting gap is fixed to a constant ($2\Delta\approx 0.12$).
In Fig.~\ref{w0 dependence}, we plot the height of the resonance peak for $|\chi_0^{(3)}(\Omega)|^2$ and 
$|\chi_{\rm vc}^{(2)}(\Omega)|^2$ as a function of the phonon frequency $\omega_0$ (with the parameters other than $\omega_0$
and $g$ the same as in Fig.~\ref{THG resolved}).
We can see that, as $\omega_0$ decreases and the effective interaction (\ref{U(w)}) becomes more retarded, 
the resonance for $\chi_{\rm vc}^{(2)}$ is enhanced while that for $\chi_0^{(3)}$ is suppressed.
This is consistent with the expectation that in the opposite antiadiabatic (non-retarded) limit ($\omega_0\to\infty$) the model approaches the attractive
Hubbard model, where the Migdal approximation is replaced by the BCS approximation, and $\chi_{\rm vc}^{(2)}$
vanishes as explained in Sec.~\ref{introduction}.
The result suggests that the retardation effect in the electron-phonon coupling indeed plays a crucial role in amplifying
the vertex correction $\chi_{\rm vc}^{(2)}$. 

We can elaborate the physical meaning of the result as follows. As we discussed previously, the dominant diagrams
contained in $\chi_{\rm vc}^{(2)}$ is the one with the resonant coupling to the light [Fig.~\ref{raman diagram}(c)].
This represents a process in which a single photon is absorbed and then emitted by electrons at different times
(with the time separation $\sim (2\Delta)^{-1}$).
The retardation effect due to the scattering of phonons (in the time scale of $\omega_0^{-1}$) can propagate between these times.
%The effect of the absorption/emission is propagated by the phonon scattering with
%the time scale of retardation $\sim\omega_0^{-1}$. 
If $2\Delta$ ($\approx 0.12$ in the present case) and $\omega_0$ are 
in the same order, 
%one cannot neglect the retardation effect. As the result in Fig.~\ref{w0 dependence} shows,
the scattering amplitude relevant for the THG resonance can be effectively enhanced,
as confirmed from the result in Fig.~\ref{w0 dependence}.
Note that the resonance between coherent phonons and the order parameter oscillation
in the regime of $\omega_0\sim 2\Delta$ has been discussed in Ref.~\onlinecite{Schnyder2011}.
%Our results suggest that one needs to seriously examine the effect of $\chi_{\rm vc}^{(2)}$,
%or in other words the contribution from the resonant diagram [Fig.~\ref{raman diagram}(c)].
%in further studies of the THG and other nonlinear optical responses of strongly correlated superconductors.

\section{Summary}
\label{summary}

To summarize, we have studied the nonlinear optical response, especially the third-harmonic generation,
for electron-phonon coupled superconductors by means of the dotted DMFT framework
proposed in the present paper. The results show that, for general polarization of the light, there is a possibility that 
the Higgs amplitude mode can contribute to the THG resonance at $2\Omega=2\Delta$
with an order of magnitude comparable to contributions from the Cooper pair breaking or charge density fluctuations, 
which is in sharp contrast to the BCS result.
The interaction between the light and Higgs mode can be mediated by the resonant coupling,
which is induced by the retarded interaction through the electron-phonon coupling.
This is confirmed by the observation that 
the intensity of the THG resonance due to the Higgs mode does indeed increase as the phonon frequency is reduced.
Let us note that the electron-phonon coupling is just one of many possibilities that could enhance the Higgs-mode effect.
These may include phonon renormalization, 
impurity scattering, dynamical correlation effects from the Coulomb interaction,
non-local correlations beyond DMFT, etc.

To make a relevance to the experiment,\cite{Matsunaga2014} it is interesting to investigate temperature dependence of the THG susceptibility.
However, we expect that this strongly depends on details of the model, since the model adopted here only includes  
a single optical phonon mode, while in realistic situations acoustic phonons may play an important role
at low temperatures. The temperature dependence may also be affected by mechanisms of energy dissipation. In this paper,
we have assumed a simple dissipation characterized by the broadening parameters $\gamma$ and $\delta$,
but it can be more complicated in real systems. Moreover, the present method has a numerical instability when
$\gamma$ or $\delta$ approaches zero.
These issues will be left as a future problem. 
%Provided 
When both the pair breaking (charge fluctuation)
and Higgs mode contribute with comparable magnitudes, as indicated here to be possible,
it is desirable to distinguish them in experiments.
One possibility is to look at the polarization dependence.\cite{Cea2016} 
To this end, one needs to accurately evaluate the polarization dependence of 
the pair breaking and Higgs-mode contributions in a realistic manner.
%such as a method based on the first principles calculation to analyze experimental data.
%The study along this direction is under way.

We wish to thank R. Shimano, R. Matsunaga, and Y. Murotani for illuminating discussions.
H.A. is supported by JSPS KAKENHI (Grant No. 26247057)  from MEXT and ImPACT project (No.~2015-PM12-05-01) from JST.
N.T. is supported by JSPS KAKENHI Grants No.~25800192 and No.~16K17729.

\appendix
\section{Notation for products of nonequilibrium Green's functions}
\label{notation}

In this appendix, we explain the notation for products of nonequilibrium Green's functions used
throughout the paper. We define
\begin{align}
[A(\omega)B(\omega')]^{R,A}
&\equiv
A^{R,A}(\omega) B^{R,A}(\omega'),
\label{AB R,A}
\\
[A(\omega)B(\omega')]^{<,>}
&\equiv
A^{R}(\omega) B^{<,>}(\omega')
+A^{<,>}(\omega) B^{A}(\omega'),
\label{AB <,>}
\end{align}
following the Langreth rule.\cite{Langreth1976} 
Here $R, A, <$, and $>$ respectively denote the retarded, advanced, lesser, and greater components
of nonequilibrium Green's functions. For the detailed definition of nonequilibrium Green's function,
we refer to Ref.~\onlinecite{noneqDMFTreview}.
The definition (\ref{AB R,A}), (\ref{AB <,>}) can be used repeatedly 
for products involving more than two Green's functions. For example, it follows 
from the above definition that
\begin{align}
[A(\omega)B(\omega')C(\omega'')]^{R,A}
&=
A^{R,A}(\omega) B^{R,A}(\omega') C^{R,A}(\omega''),
\\
[A(\omega)B(\omega')C(\omega'')]^{<,>}
&=
A^{R}(\omega) B^{R}(\omega') C^{<,>}(\omega'')
\notag
\\
&\quad
+A^{R}(\omega) B^{<,>}(\omega') C^{A}(\omega'')
\notag
\\
&\quad
+A^{<,>}(\omega) B^{A}(\omega') C^{A}(\omega'').
\end{align}
If products explicitly contain $\epsilon_{\bm k}$, we can regard it
as a component of Green's function with the definition,
\begin{align}
[\epsilon_{\bm k}]^{R,A}
&\equiv
\epsilon_{\bm k},
\\
[\epsilon_{\bm k}]^{<,>}
&\equiv
0.
\end{align}
For example, we have
\begin{align}
[\epsilon_{\bm k} A(\omega)B(\omega')]^{<,>}
&=
\epsilon_{\bm k}[A^{R}(\omega) B^{<,>}(\omega')
+A^{<,>}(\omega) B^{A}(\omega')].
\end{align}
The same applies to derivatives of $\epsilon_{\bm k}$ such as $\dot\epsilon_{\bm k}, \ddot\epsilon_{\bm k}$.

\section{A momentum integral formula for the calculation of THG}
\label{momentum integral}

In this appendix, we summarize some useful formulae for the momentum integral employed
in the calculation of the THG susceptibility in the dotted DMFT. As we have seen in Sec.~\ref{dotted dmft},
one frequently encounters a momentum integral of a function of $\epsilon_{\bm k}$ [Eq.~(\ref{ek})]
multiplied by some of $\dot\epsilon_{\bm k}, \ddot\epsilon_{\bm k}, \dddot\epsilon_{\bm k}$, and $\ddddot\epsilon_{\bm k}$
[see Eqs.~(\ref{dot ek})-(\ref{ddot ek}) for the definition]. 
In the main text, we have taken the polarization vector as
\begin{align}
\bm e
&=
\frac{1}{\sqrt{m}}(\overbrace{\underbrace{1,1,\dots,1}_m,0,\dots,0}^d).
\end{align}
In the limit $d,m\to\infty$ with a fixed ratio $\alpha=m/d$,
the momentum integral is reduced to an integral over the single variable $\epsilon=\epsilon_{\bm k}$,
as described in Sec.~\ref{model}.

Here we list the results. For momentum integrals containing two derivatives, we have
\begin{align}
\sum_{\bm k} \ddot\epsilon_{\bm k} f(\epsilon_{\bm k})
&=
-\frac{1}{d} \sum_{\bm k} \epsilon_{\bm k} f(\epsilon_{\bm k}),
\label{ddot ek f}
\\
\sum_{\bm k} (\dot\epsilon_{\bm k})^2 f(\epsilon_{\bm k})
&=
\frac{t^\ast{}^2}{d} \sum_{\bm k} f(\epsilon_{\bm k}),
\end{align}
which can be used in the calculation of $\chi_{\rm vc}^{(i)}$ [Eqs.~(\ref{chi vc1})-(\ref{chi vc2})] and the dotted lattice Dyson equation
(\ref{dotted lattice Dyson}).
For momentum integrals containing four derivatives, we have
\begin{align}
\sum_{\bm k} \ddddot\epsilon_{\bm k} f(\epsilon_{\bm k})
&=
\frac{1}{d^2\alpha}\sum_{\bm k} \epsilon_{\bm k} f(\epsilon_{\bm k}),
\\
\sum_{\bm k} \dot\epsilon_{\bm k} \dddot\epsilon_{\bm k} f(\epsilon_{\bm k})
&=
-\frac{t^\ast{}^2}{d^2\alpha}\sum_{\bm k} f(\epsilon_{\bm k}),
\\
\sum_{\bm k} (\ddot\epsilon_{\bm k})^2 f(\epsilon_{\bm k})
&=
\frac{t^\ast{}^2(1-\alpha)}{d^2\alpha}\sum_{\bm k} f(\epsilon_{\bm k})
+\frac{1}{d^2}\sum_{\bm k} \epsilon_{\bm k}^2 f(\epsilon_{\bm k}),
\\
\sum_{\bm k} (\dot\epsilon_{\bm k})^2 \ddot\epsilon_{\bm k} f(\epsilon_{\bm k})
&=
-\frac{t^\ast{}^2}{d^2}\sum_{\bm k} \epsilon_{\bm k} f(\epsilon_{\bm k}),
\\
\sum_{\bm k} (\dot\epsilon_{\bm k})^4 f(\epsilon_{\bm k})
&=
\frac{3t^\ast{}^4}{d^2}\sum_{\bm k} f(\epsilon_{\bm k}),
\end{align}
which can be used in the calculation of $\chi_0^{(i)}$ [Eqs.~(\ref{chi0 1})-(\ref{chi0 5})].

Once the momentum integral is reduced to an integral over $\epsilon=\epsilon_{\bm k}$,
it can be further evaluated analytically. To see this, let us parametrize the self-energy
(at half filling with $\mu=0$) as
\begin{align}
\hat\Sigma^R(\omega)
&=
[1-Z(\omega)]\omega+\phi(\omega)\tau_1.
\end{align}
We define functions
\begin{align}
\hat S^\pm(\omega)
&\equiv
Z(\omega)\omega\pm \phi(\omega)\tau_1,
\\
S^2(\omega)
&\equiv
[Z(\omega)\omega]^2-\phi(\omega)^2,
\end{align}
with which the retarded Green's function is represented as
\begin{align}
\hat G_{\bm k}^R(\omega)
&=
[\omega-\epsilon_{\bm k}\tau_3-\hat\Sigma^R(\omega)]^{-1}
=
\frac{\hat S^+(\omega)+\epsilon_{\bm k}\tau_3}{S^2(\omega)-\epsilon_{\bm k}^2}.
\end{align}
As an example, let us consider the first term in the dotted lattice Dyson equation (\ref{dotted lattice Dyson}),
whose retarded component can be evaluated with Eq.~(\ref{ddot ek f}) as
\begin{align}
&
\sum_{\bm k} \hat G_{\bm k}^R(\omega+2\Omega)\ddot\epsilon_{\bm k}\tau_3 \hat G_{\bm k}^R(\omega)
\notag
\\
&=
-\frac{1}{d}\sum_{\bm k} 
\frac{\hat S^+(\omega+2\Omega)+\epsilon_{\bm k}\tau_3}{S^2(\omega+2\Omega)-\epsilon_{\bm k}^2}
\epsilon_{\bm k}\tau_3
\frac{\hat S^+(\omega)+\epsilon_{\bm k}\tau_3}{S^2(\omega)-\epsilon_{\bm k}^2}
\notag
\\
&=
\frac{1}{d}
\frac{\hat G^R(\omega)\hat S^-(\omega)-\hat G^R(\omega+2\Omega)\hat S^-(\omega+2\Omega)}{S^2(\omega)-S^2(\omega+2\Omega)}
\notag
\\
&\quad\times
[\hat S^+(\omega)+\hat S^+(\omega+2\Omega)].
\end{align}
Note that $\hat G^R(\omega)$ and $\hat S^\pm(\omega')$ commute with each other.
In this way, every momentum integral appearing in the calculation of the dotted DMFT
and nonlinear optical susceptibilities can be written in terms of the local Green's function and self-energy.
This greatly reduces the computational cost of the dotted DMFT algorithm.

\bibliographystyle{apsrev}
\bibliography{ref}

\end{document}